\documentclass[twocolumn]{IEEEtran}
\usepackage{cite}
\usepackage{amsmath}
\usepackage{mathtools}
\usepackage{amssymb}
\usepackage[cmintegrals]{newtxmath}
\usepackage{algorithm,algorithmic}
\usepackage{array}

\usepackage[english]{babel}

\begin{document}

\title{Interference Exploitation in Full Duplex Communications: Trading Interference Power for Both Uplink and Downlink Power Savings}

\author{Mahmoud~T.~Kabir,~\IEEEmembership{Student Member,~IEEE,}
       Muhammad~R.~A.~Khandaker,~\IEEEmembership{Member,~IEEE,}
       and~Christos~Masouros,~\IEEEmembership{Senior Member,~IEEE}} 
\maketitle

\begin{abstract}
This paper considers a multiuser full-duplex (FD) wireless communication system, where a FD radio base station (BS) serves multiple single-antenna half-duplex (HD) uplink and downlink users simultaneously. Unlike conventional interference mitigation approaches, we propose to use the knowledge of the data symbols and the channel state information (CSI) at the FD radio BS to exploit the multi-user interference constructively rather than to suppress it. We propose a multi-objective optimisation problem (MOOP) via the weighted Tchebycheff method to study the trade-off between the two desirable system design objectives namely the total downlink transmit power minimisation and the total uplink transmit power minimisation problems at the same time ensuring the required quality-of-service (QoS) for all users. In the proposed MOOP, we adapt the QoS constraints for the downlink users to accommodate constructive interference (CI) for both generic phase shift keying (PSK) modulated signals as well as for quadrature amplitude modulated (QAM) signals. We also extended our work to a robust design to study the system with imperfect uplink, downlink and self-interference CSI. Simulation results and analysis show that, significant power savings can be obtained. More importantly, however, the MOOP approach here allows for the power saved to be traded off for both uplink and downlink power savings, leading to an overall energy efficiency improvement in the wireless link.               
\end{abstract}


\section{Introduction}
The ever-increasing need for improved spectrum-efficiency in wireless links has brought FD at the forefront of research attention. By allowing simultaneous transmission and reception, FD since it has the potential to drastically improve the spectral efficiency of the HD communication networks \cite{bharadia2013full,choi2010achieving,nguyen2014spectral,song2015resource,ng2012dynamic,ngo2014multipair}. One major hurdle with the FD communication systems is the self-interference (SI) from the transmit antennas to the receive antennas of the wireless transceiver. This interference raises the noise floor and it becomes a dominant factor in the performance of the FD system. However, major breakthroughs have been made in practical FD system setups \cite{bharadia2013full} and \cite{choi2010achieving} that show that the SI can be partially cancelled to within a few dB of the noise floor. While others focused on resource management, in \cite{nguyen2014spectral}, the authors investigated the spectral efficiency of FD small cell wireless systems by considering a joint beamformer design to maximize the spectral efficiency subject to power constraints. In \cite{song2015resource}, the authors discussed the resource allocation problems in FD-MIMO, FD-Relay, FD-OFDMA and FD-HetNet systems including power control, interference-aware beamforming, e.t.c. Also, resource allocation and scheduling in FD-MIMO-OFDMA relaying systems was studied in \cite{ng2012dynamic}. In \cite{ngo2014multipair}, the authors used massive arrays at the FD relay station to cancel out loop interference and as a result increase the sum spectral efficiency of the system.  \par 
Many of the above FD solutions build upon existing beamforming solutions in the literature, that have been extensively developed for the downlink channel, moving from the sophisticated but capacity achieving non-linear beamforming techniques \cite{costa1983writing,erez2005capacity,masouros2012interference,windpassinger2004precoding,garcia2014power} to the less complex linear beamforming techniques \cite{peel2005vector,masouros2009dynamic,masouros2011correlation,alsusa2008adaptive,masouros2014maximizing}. Several optimization based schemes that provide optimal solutions subject to required quality of service (QoS) constraints have been proposed for multi-input single-output (MISO) systems in \cite{bengtsson2001handbook,bengtsson1999optimal,rashid1998transmit,visotsky1999optimum}. In \cite{vucic2009robust,zheng2008robust}, the authors addressed the problem of robust designs in downlink multiuser MISO systems with respect to erroneous channel state information (CSI). The work in \cite{schubert2004solution} focused on addressing both max-min signal-to-interference (SINR) balancing problem and power minimisation problem with SINR constraints. More recently, it has been shown in \cite{masouros2009dynamic,masouros2011correlation,masouros2013known,zheng2014rethinking} that with the knowledge of the users' data symbols and the CSI, the interference can be classified into constructive and destructive interference. And further findings in \cite{alodeh2013data,alodeh2014multicast,alodeh2015constructive,alodeh2015energy,alodeh2016energy,alodeh2015constructive2,masouros2015exploiting,law2016constructive,li2017exploiting,amadori2016constant,Law:inpress-a,Law2:inpress-a} show that tremendous gains can be achieved by exploiting the constructive interference based on symbol level optimization for both PSK and QAM modulations. However, these findings are all based on MISO HD systems.  \par
Our work extends the above interference exploitation concept to the FD transmission, by employing multi-objective optimization, as most recently studied for FD in \cite{sun2015multi,sun2016multi,leng2016multi}. The authors in \cite{sun2015multi} investigated the power efficient resource allocation for a MU-MIMO FD system. They proposed a multi-objective optimisation problem (MOOP) to study the total uplink and downlink transmit power minimization problems jointly via the weighed Tchebycheff method. They extended their work to a robust and secure FD systems model in the presence of roaming users (eavesdroppers) in \cite{sun2016multi}. Similarly, in \cite{leng2016multi} the authors used a similar model to investigate the resource allocation for FD simultaneous wireless information and power transfer (SWIPT) systems. Accordingly, in this work we aim to further reduce the power consumption in FD MU-MIMO wireless communication systems by adopting the concept of constructive interference in the literature to the downlink channel for both PSK and QAM modulation. By exploiting interference constructively, useful signal power from interference, we can provide a truly power efficient resource allocation for a FD MU-MIMO system. The interference exploitation concept is yet to be explored in the realm of FD transmission, where FD offers the unique opportunity to trade-off the harvested interference power for both uplink and downlink power savings through the MOOP designs. Against the state-of-the-art, we summarize our contributions below:
\begin{enumerate}
    \item We first formulate the FD beamforming design problem that minimizes (a)the total downlink transmit power and, (b)the total uplink transmit power problem, for PSK and QAM modulation separately. Both problems are subject to downlink users SINR requirement based on the constructive interference regions and uplink users SINR requirement. Unlike conventional FD beamformers, we show that the proposed optimizations are convex and can be easily solved by conventional solvers.    
    \item Building on the above single-objective problems, we then formulate a multi-objective problem to study the trade-off between the total uplink and downlink transmit power minimization problems jointly via the weighed Tchebycheff method. Again, unlike the conventional FD beamformers, we show that the proposed optimization is convex.  
    \item We further derive robust MOOP for both the conventional and the proposed interference exploitation approach by recasting the MOOP into a virtual multicast problem for erroneous downlink, uplink and SI CSI with bounded errors.
\end{enumerate}
\par
The rest of the paper is organised as follows. Section II introduces the system model that is considered in this paper. Section III describes the two conventional power minimisation problems of interest to the system operator and then briefly describes the MOOP formulation based on the two problems. In Section IV, the proposed power minimization optimisation problems based on constructive interference regions are presented for PSK and QAM modulations. Then in Section V, we present the robust version of the optimisation problem presented in Section IV. In Section VI, we provide a computational complexity analysis of the MOOP formulations. Section VII illustrates the important results and discussions. And finally we conclude in Section VIII.

\section{System Model}
We consider a FD multiuser communication system as shown in Fig. 1. The system consists of a FD radio BS with {\textit{N}}  antennas serving {\textit{K}} HD downlink users and {\textit{J}}  HD uplink users. Each user is equipped with a single antenna to reduce hardware complexity. Let \( {\textbf{h}_i}  \in \mathbb{C}^{N \times 1} \)   be the channel vector between the FD radio BS and the {\textit{i-th}} downlink user, and \( {\textbf{f}_j} \in \mathbb{C}^{N \times 1} \) be the channel vector between the FD radio BS and the {\textit{j-th}} uplink user. We denote the transmit signal vector from the FD radio BS to the {\textit{i-th}} downlink user as
\begin{align}
{\textbf{t}_i} &= {\textbf{w}_i}d_i        
\end{align}
where \({\textbf{w}_i} \in \mathbb{C}^{N \times 1} \) and \( d_i \) denote the beamforming vector and the unit data symbol for the {\textit{i-th}} downlink user. The received signal at the {\textit{i-th}} downlink user is:
\begin{align}
y_i &= \underset{\textrm{desired signal}}{\underbrace{{\textbf{h}_i^H}{\textbf{t}_i}}}+ \underset{\textrm{interference plus noise}}{\underbrace{\sum_{k\ne i}^{K} {\textbf{h}_i^H}{\textbf{t}_k} + n_i}} 
\end{align}
where \( n_i \sim {\mathcal{CN}} \left(0,\sigma_i ^2 \right)  \) represents the additive white Gaussian noise AWGN at the \({\textit{i-th}} \) downlink user. For each time slot the FD radio BS transmits {\textit{K}} independent unit data symbols {\textit{d}} simultaneously at the same frequency to the {\textit{K}} downlink users. The first term in (2) represents the desired signal while the second term is the multiuser interference signal. The received signal from the {\textit{J}} uplink users at the FD radio BS is:
\begin{figure}[t]
\centering
\includegraphics[width=8cm]{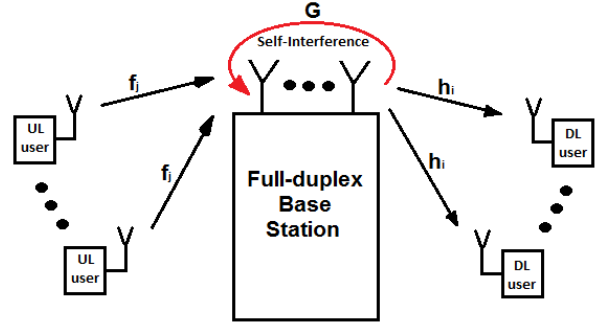}  
\caption{System model with a FD radio BS with {\textit{N}} antennas, {\textit{K}} HD downlink users and {\textit{J}} HD uplink users.} 
\end{figure}
\begin{align}
{\textbf{y}^{BS}} &= \sum_{j = 1}^{J} {\sqrt P_j}{\textbf{f}_j}x_j + \underset{\textrm{residual self-interference}}{\underbrace{{\textbf{G}}\sum_{k = 1}^{K} {\textbf{t}_k}}} + {\textbf{z}} 
\end{align}
where \({\textit{P}_j}\) and \({\textit{x}_j}\) denotes the uplink transmit power and the data symbol from the {\textit{j-th}} uplink user respectively. The vector  \( {\textbf{z}}  \sim {\mathcal{CN}} (0, {\sigma}_N^2)\) represents the additive white Gaussian noise AWGN at the FD radio BS. The matrix \( {\textbf{G}} \in \mathbb{C}^{N \times N} \) denotes the self-interference (SI) channel at the FD radio BS. In the literature, different SI mitigation techniques have been proposed \cite{bharadia2014full,day2012full} to reduce the effect of self-interference. In order to isolate our proposed scheme from the specific implementation of a SI mitigation technique, since the SI cannot be cancelled perfectly in FD systems due to limited dynamic range at the receiver even if the SI channel is known perfectly \cite{sun2016multi,day2012full}, we model the residual SI after cancellation as \( \left({\textbf{G}}\sum_{k = 1}^{K} {\textbf{t}_k} \right) \) as in \cite{sun2015multi,sun2016multi}. Accordingly, the first term of (3) represents the desired signal from the {\textit{j-th}} uplink user and the second term represents the residual SI. 
\par Before we formulate the problem, we first define the signal-to-interference ratio (SINR) at the {\textit{i-th}} downlink user and at the FD radio BS respectively as
\begin{align}
SINR_i^{DL} &=  \frac {{\mid{\textbf{h}_i^H}{\textbf{w}_i}\mid}^2 }{\sum_{k\ne i}^{K} \mid{\textbf{h}_i^H}{\textbf{w}_k}\mid ^2 + \sigma_i ^2 } 
\end{align}
\begin{align}
SINR_j^{UL} &=  \frac {{P_j \mid{\textbf{f}_j^H}{\textbf{u}_j}\mid}^2} {\sum_{n\ne j}^{J}  P_n \mid{\textbf{f}_n^H}{\textbf{u}_j}\mid ^2 + \sum_{k= 1}^{K} \mid{\textbf{u}_j^H}{\textbf{G}}{\textbf{w}_k}\mid ^2 + \sigma_N ^2 {\lVert {\textbf{u}_j} \rVert}^2} 
\end{align}
\\
where  \( {\textbf{u}_j}  \in ^{N \times 1} \) is the receive beamforming vector for detecting the receivied symbol from the {\textit{j-th}} uplink user. To reduce complexity, we assume a zero-forcing receiver at the BS. Hence, the receive beamformer for the {\textit{j-th}} uplink user is given as
\begin{align}
{\textbf{u}_j} &= ({\textbf{r}_j}{\textbf{F}^\dagger})^H   
\end{align}
where \( {\textbf{r}_j} = [ \underset{j-1}{\underbrace{ 0,\hdots,0,} }1,\underset{J-j}{\underbrace{ 0,\hdots,0}} ] \), \({\textbf{F}^\dagger} = ({\textbf{F}}^H{\textbf{F}})^{-1}{\textbf{F}}^H, ^\dagger \) denotes the pseudo-inverse operation and \( {\textbf{F}} = [{\textbf{f}}_1,\dots,{\textbf{f}}_J ] \).

\section{Conventional Power Minimization Problem}
In this section, we study the conventional power minimization (PM) problem where all the interferences are treated as undesired signals. We first formulate the downlink and uplink power minimization problems, which aim to minimize the total average downlink and uplink transmit power, respectively, subject to the downlink users SINR and uplink users SINR. Then we formulate a multi-objective PM problem that aims to investigate the two system's objectives (downlink and uplink) jointly.   

\subsection*{Problem 1: Total Downlink Transmit PM Problem }
The downlink PM problem for FD optimisation is typically formulated as \cite{sun2015multi,sun2016multi}:
\begin{equation} 
\begin{split}
\mathcal{P}1: \quad  \underset{{\textbf{w}_i},P_j}{\text{min}} \quad & \sum_{i=1}^{K} {\left\|{\textbf{w}_i}\right\|}^2 \\
\text{s.t.} \quad & A1: \frac {{\mid{\textbf{h}_i^H}{\textbf{w}_i}\mid}^2 }{\sum_{k\ne i}^{K} \mid{\textbf{h}_i^H}{\textbf{w}_k}\mid ^2 + \sigma_i ^2 }  \geq \Gamma_i^{DL}, \forall i,\\
& A2: \frac {{P_j \mid{\textbf{f}_j^H}{\textbf{u}_j}\mid}^2} {I_j + \sigma_N ^2 {\lVert {\textbf{u}_j} \rVert}^2} \geq \Gamma_j^{UL}, \forall j 
\end{split}
\end{equation}

where, \(I_j = \sum_{n\ne j}^{J} P_n \mid{\textbf{f}_n^H}{\textbf{u}_j}\mid ^2 + \sum_{k= 1}^{K} \mid{\textbf{u}_j^H}{\textbf{G}}{\textbf{w}_k}\mid ^2  \), we define \( \Gamma_i^{DL} \) and \( \Gamma_j^{UL} \) as the minimum required SINRs for the {\textit{i-th}} downlink user and the {\textit{j-th}} uplink user, respectively. This problem aims to minimize the total downlink transmit power with no regards to the consumed uplink transmit power. This problem is non-convex and it is commonly solved via semidefinite relaxation as in \cite{sun2015multi,sun2016multi}.  
\subsection*{Problem 2: Total Uplink Transmit PM Problem } 
The uplink PM problem for FD optimisation is typically formulated as \cite{sun2015multi,sun2016multi}:
\begin{equation} 
\begin{split}
\mathcal{P}2: \quad  \underset{{\textbf{w}_i},P_j}{\text{min}} \quad & \sum_{j=1}^{J} {P_j} \\
\text{s.t.} \quad & A1: \frac {{\mid{\textbf{h}_i^H}{\textbf{w}_i}\mid}^2 }{\sum_{k\ne i}^{K} \mid{\textbf{h}_i^H}{\textbf{w}_k}\mid ^2 + \sigma_i ^2 }  \geq \Gamma_i^{DL}, \forall i, \\
& A2: \frac {{P_j \mid{\textbf{f}_j^H}{\textbf{u}_j}\mid}^2} {I_j + \sigma_N ^2 {\lVert {\textbf{u}_j} \rVert}^2} \geq \Gamma_j^{UL}, \forall j 
\end{split}
\end{equation}  
where, \( \Gamma_i^{DL} \) and \( \Gamma_j^{UL} \) are the minimum required SINRs for the {\textit{i-th}} downlink user and the {\textit{j-th}} uplink user, respectively. This problem unlike problem \( \mathcal{P}1 \) aims to minimize the total uplink transmit power with no regards to the consumed downlink transmit power. Problem \(\mathcal{P}2 \) is non-convex and it is commonly solved via semidefinite relaxation as in \cite{sun2015multi,sun2016multi}.

\subsection*{Problem 3: Multi-objective PM Problem}
This formulation combines the two objectives of problem \( \mathcal{P}1 \) and \( \mathcal{P}2 \) since both objectives are very important to both the users and system operator. The multi-objective optimization is employed when there is need to study jointly the trade-off between two desirable objectives via the concept of Pareto optimality. A point is said to be Pareto optimal if there is no other point that improves any of the objectives without decreasing the others \cite{marler2004survey}. \cite{marler2004survey} did a survey of multi-objective optimization methods in engineering. By using the weighted Tchebycheff method \cite{marler2004survey} which can acheive the complete Pareto optimal set with lower computational complexity, the multi-objective PM problem for FD optimisation is typically formulated as \cite{sun2015multi,sun2016multi},
\begin{equation} 
\begin{split}
\mathcal{P}3: \quad  \underset{{\textbf{w}_i},P_j}{\text{min}} \quad & \underset{a=1,2}{\text{max}}  \left\{{\lambda_{a}}\left(R_a^* - R_a({\textbf{w}_i},P_j)\right)\right\} \\
\text{s.t.} \quad & A1: \frac {{\mid{\textbf{h}_i^H}{\textbf{w}_i}\mid}^2 }{\sum_{k\ne i}^{K} \mid{\textbf{h}_i^H}{\textbf{w}_k}\mid ^2 + \sigma_i ^2 }  \geq \Gamma_i^{DL}, \forall i, \\
& A2: \frac {{P_j \mid{\textbf{f}_j^H}{\textbf{u}_j}\mid}^2} {I_j + \sigma_N ^2 {\lVert {\textbf{u}_j} \rVert}^2} \geq \Gamma_j^{UL}, \forall j 
\end{split}
\end{equation}

where \( R_a \) and \( R_a^* \) denote the objective value and the optimal objective value of the \( {\textit{a-th}} \) optimisation problem, respectively. The variable \( {\lambda_{a}} \geq 0 \), \( \sum {\lambda_{a}} = 1 \), specifies the priority given to the \( {\textit{a-th}} \) objective i.e. for a given \( {\lambda_{1}} = 0.8  \) means \( 80\% \) priority is given to the objective of problem \( \mathcal{P}1 \) and  \( 20\% \) priority to the objective of problem \( \mathcal{P}2 \). By varying  \( {\lambda_{a}} \) we can obtain the complete Pareto optimal set. Problem \( \mathcal{P}3 \) is a non-convex problem due to the SINR constraints A1 and A2, and it is commonly solved via semidefinite relaxation as in \cite{sun2015multi,sun2016multi}.    

\setcounter{equation}{15}
\begin{figure*}[!b]  
\vspace*{4pt}
\hrulefill
\begin{equation} 
\begin{split}
\mathcal{P}4: \quad \underset{{\textbf{w}_k},{P_j}}{\text{min}} \quad & {\left\| \sum_{k=1}^{K} {\textbf{w}_k}{e^{j({\phi}_k - {\phi}_1)}}\right\|}^2 \\
\text{s.t.} \quad & B1: \left| Im \left({\textbf{h}_i^H} \sum_{k= 1}^{K} {\textbf{w}_k}{e^{j({\phi}_k - {\phi}_i)}} \right) \right| \leq \left(Re \left({\textbf{h}_i^H} \sum_{k= 1}^{K} {\textbf{w}_k}{e^{j({\phi}_k - {\phi}_i)}}\right) - {\sqrt{\Gamma_i^{DL} \sigma_i^2} }\right) \tan \theta, \forall i, \\
& B2: \frac {{P_j \left|{\textbf{f}_j^H}{\textbf{u}_j}\right|}^2} {\sum_{n\ne j}^{J} P_n \left|{\textbf{f}_n^H}{\textbf{u}_j}\right| ^2 + \sum_{k= 1}^{K} \left|{\textbf{u}_j^H}{\textbf{G}}{\textbf{w}_k{e^{j({\phi}_k - {\phi}_1)}}}\right| ^2 + \sigma_N ^2 {\lVert {\textbf{u}_j} \rVert}^2}   \geq \Gamma_j^{UL}, \forall j 
\end{split}
\end{equation} 
\end{figure*}
\section{Power Minimization Problem based on Constructive Interference}
In this section, we study the PM optimization problems based on constructive interference. With prior knowledge of the CSI and users' data symbols for the downlink users, the instantaneous interference can be exploited rather than suppressed \cite{masouros2015exploiting}. To be precise, constructive interference is the interference that pushes the received signal further into the detection region of the constellation and away from the detection threshold \cite{masouros2015exploiting}. This concept has been thoroughly studied in the literature for both PSK and QAM modulation. We refer the reader to \cite{alodeh2013data,alodeh2014multicast,alodeh2015constructive,alodeh2015energy,alodeh2016energy,alodeh2015constructive2,masouros2015exploiting,law2016constructive} for further details of this topic. Motivated by this idea, here, we apply this concept to the PM problems in Section III for both PSK and QAM modulations. We note that constructive interference is only applied to the downlink users and not the uplink users following that only the prior knowledge of the CSI and users' data symbols for the downlink users are available at the BS. Nevertheless, we show in the following that power savings can be obtained for both uplink and downlink transmission, by means of the MOOP design.

\subsection{Constructive Interference for PSK modulation}
To illustrate this concept, we provide a geometric illustration of the constructive interference regions for a QPSK constellation in Fig. 2. We can define the total transmit signal vector as 
\setcounter{equation}{9}
\begin{align}
\sum_{k = 1}^{K}{\textbf{w}_k}d_k  = \sum_{k=1}^{K} {\textbf{w}_k}{e^{j({\phi}_k - {\phi}_i)}}d_i     
\end{align}

where \( d_i = de^{{\phi}_i} \) is the desired symbol for the {\textit{i-th}} downlink user. Therefore, the received signal (2) at the {\textit{i-th}} downlink user can be redefined as
\begin{align}
y_i &= {\textbf{h}_i^H}\sum_{k = 1}^{K} {\textbf{w}_k}d_k + m_i \\
	 &= {\textbf{h}_i^H}\sum_{k = 1}^{K} {\textbf{w}_k}{e^{j({\phi}_k - {\phi}_i)}}d_i + m_i
\end{align} \par
Accordingly, since the interference contributes constructively to the received signal, it has been shown in \cite{masouros2011correlation} that the downlink SNR at the {\textit{i-th}} downlink user (4) can be rewritten as
\begin{align}
SNR_i^{DL} &=  \frac {{\left|{\textbf{h}_i^H}\sum_{k = 1}^{K} {\textbf{w}_k}d_k\right|}^2 }{\sigma_i ^2 } 
\end{align} \par
Without loss of generality, by taking user 1 as reference the instantaneous transmit power for a unit symbol is 
\begin{align}
P_{total} = {\left\| \sum_{k=1}^{K} {\textbf{w}_k}{e^{j({\phi}_k - {\phi}_1)}} \right\|}^2
\end{align}
\par
As detailed in \cite{masouros2015exploiting}, the shaded area in Fig. 2 is the region of constructive interference. If the recieved signal \( y_i \) falls within this region, then interference has been exploited constructively. The angle \( \theta = \pm\frac{\pi}{M} \) determines the maximum angle shift of the constructive interference region for a modulation order \( M \), \( a_I \) and \( a_R \) are the imaginary and real parts of the received signal \( y_i \) without the noise, respectively. The detection threshold is determined by \( \gamma=\sqrt{\Gamma_i^{DL} \sigma_i} \). \par
	Therefore, by applying these definitions and basic geometry from Fig. 2 it can be seen that for the received signal to fall in the constructive region of the constellation we need to have \( a_I \leq (a_R - \gamma)\tan \theta  \). Accordingly, we can define the downlink SINR constraint that guarantees constructive interference at the {\textit{i-th}} downlink user by
\begin{multline}
\left| Im \left({\textbf{h}_i^H} \sum_{k= 1}^{K} {\textbf{w}_k}{e^{j({\phi}_k - {\phi}_i)}} \right) \right| \leq \\
 		\left(Re \left({\textbf{h}_i^H} \sum_{k= 1}^{K} {\textbf{w}_k}{e^{j({\phi}_k - {\phi}_i)}}\right) - {\sqrt{\Gamma_i^{DL} \sigma_i^2} }\right) \tan \theta
\end{multline}   
\begin{figure}[t]
\centering
\includegraphics[width=8cm]{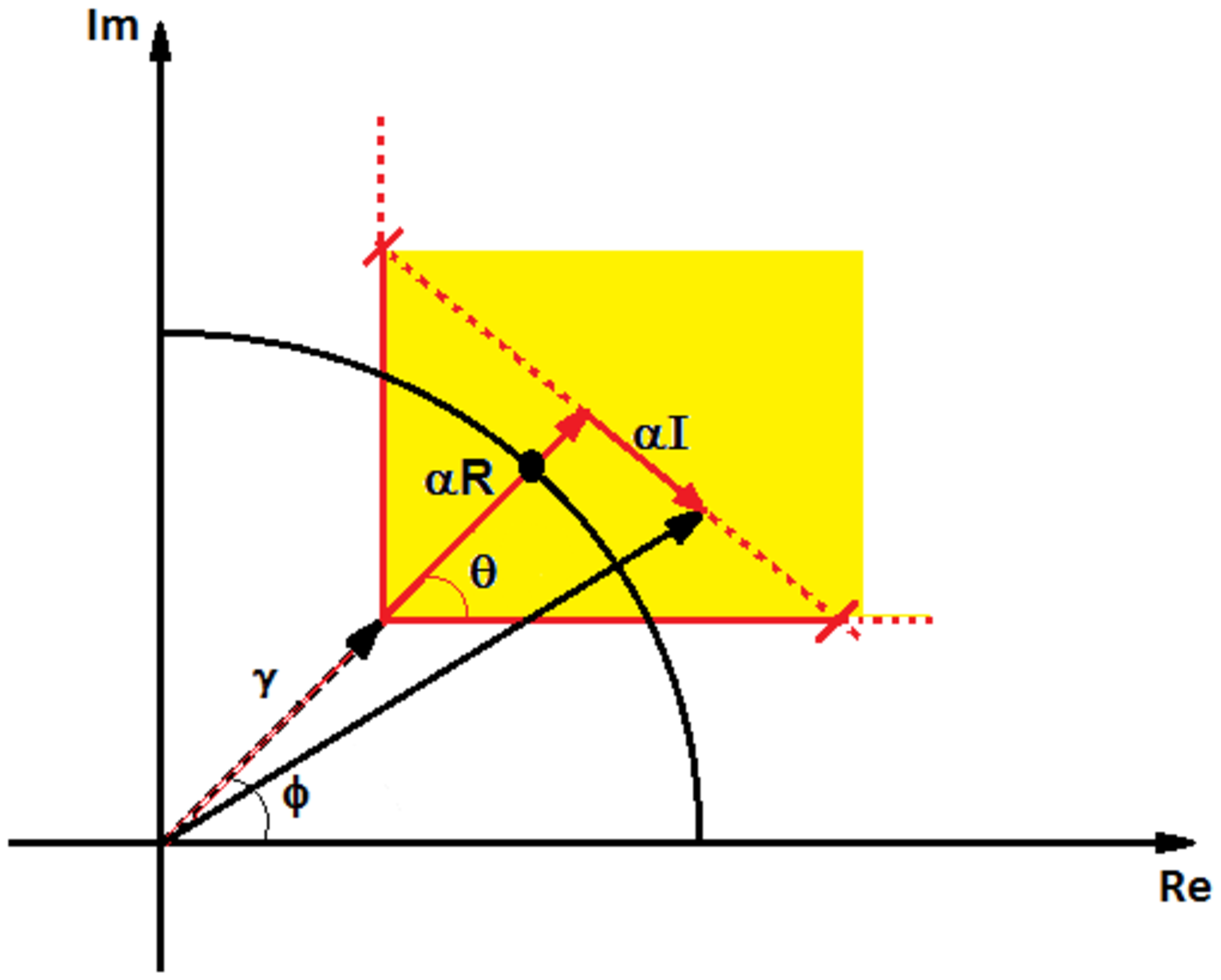}
\caption{Constructive interference region for a QPSK constellation point} 
\end{figure}

\subsubsection{Total Downlink Transmit PM Problem}
Based on the analysis above, we can modify the SINR constraints for the downlink users to accommodate CI. The optimisation problem for the total downlink transmit PM is expressed in \( \mathcal{P}4 \). The total minimum downlink transmit power is minimised subject to constraint B1, which guarantees constructive interference for the downlink users for minimum required SINR \( \Gamma_i^{DL} \) while the constraint B2 guarantees that the uplink users their minimum required SINR \( \Gamma_j^{UL} \). Unlike its conventional counterpart \( \mathcal{P}1 \), it can be seen that \( \mathcal{P}4 \) is convex due to the substitution of the conventional downlink SINR constraint with the CI SNR constraints and can be tackled with standard solvers.     

\subsubsection{Total Uplink Transmit PM Problem} 
On the other hand, we formulate the uplink transmit PM problem by minimising the total uplink transmit power with no regards to the downlink transmit power.
\setcounter{equation}{16}
\begin{equation} 
\begin{split}
\mathcal{P}5: \, \underset{{\textbf{w}_i},P_j}{\text{min}} \,  & \sum_{j=1}^{J} {P_j} \\
\text{s.t.} \quad & B1: \left| Im \left({\textbf{h}_i^H} \sum_{k= 1}^{K} {\textbf{w}_k}{e^{j({\phi}_k - {\phi}_i)}} \right) \right| \\
& \leq \left(Re \left({\textbf{h}_i^H} \sum_{k= 1}^{K} {\textbf{w}_k}{e^{j({\phi}_k - {\phi}_i)}}\right) - {\sqrt{\Gamma_i^{DL} \sigma_i^2} }\right) \tan \theta, \forall i, \\
& B2: \frac {{P_j \left|{\textbf{f}_j^H}{\textbf{u}_j}\right|}^2} {I^{PSK}_j + \sigma_N ^2 {\lVert {\textbf{u}_j} \rVert}^2} \geq \Gamma_j^{UL}, \forall j 
\end{split}
\end{equation}
where, \( I^{PSK}_j = \sum_{n\ne j}^{J} P_n \left|{\textbf{f}_n^H}{\textbf{u}_j}\right| ^2 + \sum_{k= 1}^{K} \left|{\textbf{u}_j^H}{\textbf{G}}{\textbf{w}_k{e^{j({\phi}_k - {\phi}_1)}}}\right| ^2 \).

Again, it can be seen that the above problem is convex and can be tackled with standard solvers.

\subsubsection{Multi-objective PM Problem}
By adapting the downlink SINR constraints in \( \mathcal{P}2 \), we can further obtain the MOOP for interference exploitation in the FD scenario under study as
\begin{equation} 
\begin{split}
\mathcal{P}6: \, \underset{{\textbf{w}_i},P_j,t}{\text{min}} \quad & t \\
\text{s.t.} \quad & B1: \left| Im \left({\textbf{h}_i^H} \sum_{k= 1}^{K} {\textbf{w}_k}{e^{j({\phi}_k - {\phi}_i)}} \right) \right| \\
& \leq \left(Re \left({\textbf{h}_i^H} \sum_{k= 1}^{K} {\textbf{w}_k}{e^{j({\phi}_k - {\phi}_i)}}\right) - {\sqrt{\Gamma_i^{DL} \sigma_i^2} }\right) \tan \theta, \forall i, \\
& B2: \frac {{P_j \left|{\textbf{f}_j^H}{\textbf{u}_j}\right|}^2} {I^{PSK}_j + \sigma_N ^2 {\lVert {\textbf{u}_j} \rVert}^2} \geq \Gamma_j^{UL}, \forall j , \\
& B3: {\lambda_{a}}\left(R_a^* - R_a({\textbf{w}_i},P_j)\right) \leq t, \forall a\in\left\{1,2\right\}.
\end{split}
\end{equation}

where \( t \) is an auxiliary variable. \par 
It can be observed that, due to the substitution of the conventional downlink SINR constraint with the CI SNR constraints, this formulation unlike the conventional problem in \( \mathcal{P}3 \) is convex and thus can be optimally solved using standard convex softwares like CVX \cite{grant2008cvx}.  

\subsection{Constructive Interference for QAM modulation}

\begin{figure}[t]
\centering
\includegraphics[width=8cm]{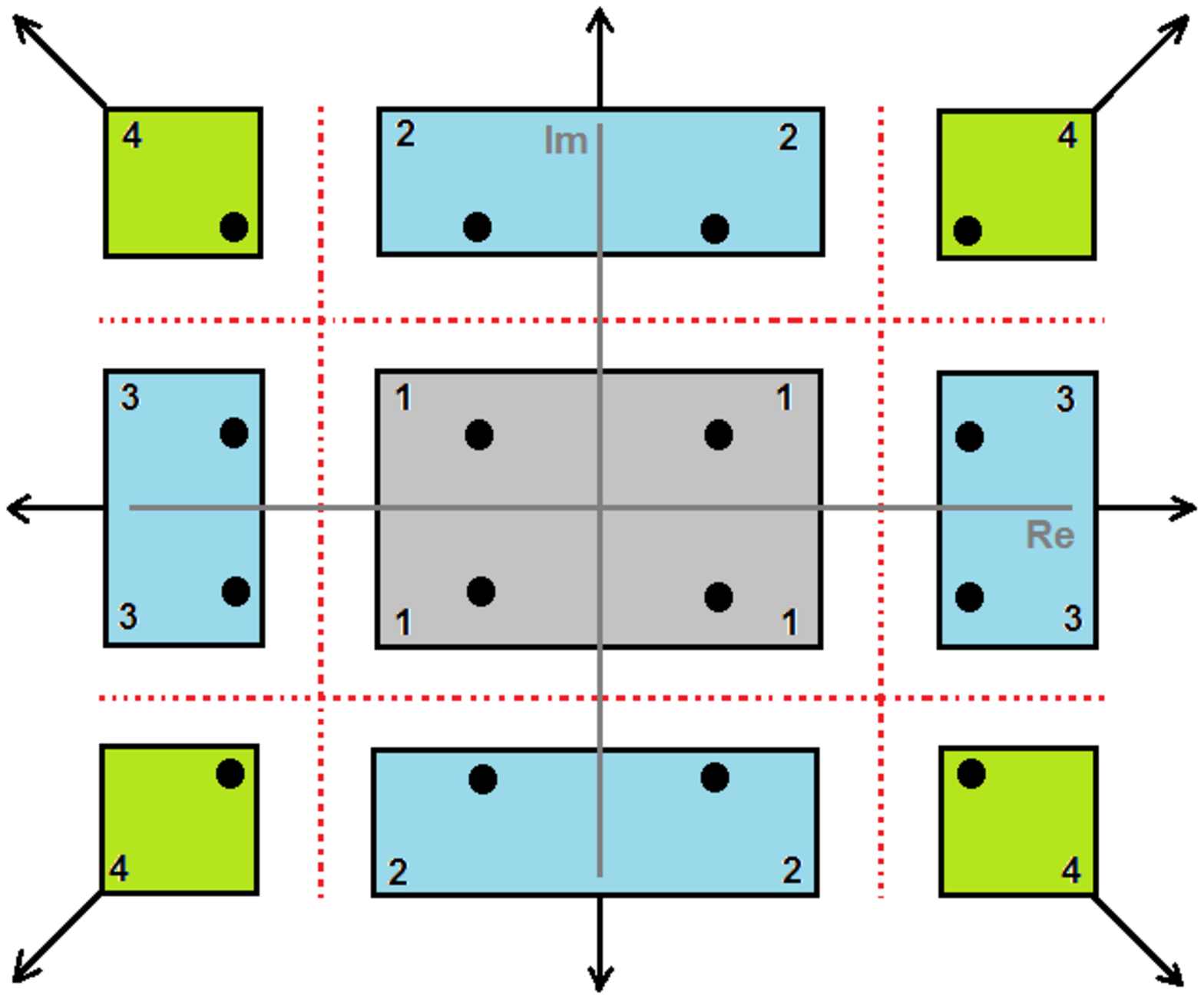}  
\caption{Schematic representation of 16QAM constellation points} 
\end{figure}

To illustrate the concept of constructive interference for QAM modulation we provide a schematic representation for 16QAM constellation points in Fig. 3. Based on \cite{alodeh2015constructive2}, to guarantee constructive interference for the constellation points, we rewrite the SINR constraints for the downlink users to exploit the specific detection regions for each group of constellation points separately as detailed below. First, we redefine the received signal without noise (12) and the instantaneous transmit power (14) in terms of amplitude not phase as
\begin{align}
y_i = {\textbf{h}_i^H} \sum_{k= 1}^{K} {\textbf{w}_k}d_k, \forall i
\end{align}    
and,
\begin{align}
P_{total} = {\left\| \sum_{k=1}^{K} {\textbf{w}_k}d_k \right\|}^2   
\end{align}
From Fig. 3, to ensure constructive interference is achieved and the constellation points are received in the correct detection region for the downlink users, the following constraints are adopted. Note that the dotted lines in Fig. 3 represent the decision boundaries.

\begin{itemize}
\item For the group of constellation points in the box labelled "1" in Fig. 3, since they are all surrounded by the decision boundaries, the constraints should guarantee that the received signals achieve the exact constellation point so as not to exceed the decision boundaries. The constraints are
\begin{align*}
C1: & Re(y_i) = {\sqrt{\Gamma_i^{DL}}}\sigma_iRe(d_i) 
\end{align*}
\begin{align*}
C2: & Im(y_i) = {\sqrt{\Gamma_i^{DL}}}\sigma_iIm(d_i) 
\end{align*}

\item For the group of constellation points labelled "2" in Fig. 3, the constraints should guarantee that the received signals fall in the detection region away from the decision boundaries, which is the real axis. The constraints are
\begin{align*}
C1: & Re(y_i) = {\sqrt{\Gamma_i^{DL}}}\sigma_iRe(d_i) 
\end{align*}
\begin{align*}
C2: & Im(y_i) \gtreqless {\sqrt{\Gamma_i^{DL}}}\sigma_iIm(d_i) 
\end{align*}

\item For the group of constellation points labelled "3" in Fig. 3, the constraints should guarantee that the received signals fall in the detection region away from the decision boundaries, which is the imaginary axis. The constraints are 
\begin{align*}
C1: & Re(y_i) \gtreqless {\sqrt{\Gamma_i^{DL}}}\sigma_iRe(d_i) 
\end{align*}
\begin{align*}
C2: & Im(y_i) = {\sqrt{\Gamma_i^{DL}}}\sigma_iIm(d_i) 
\end{align*}

\item For the group of constellation points labelled "4" in Fig. 3, the constraints should guarantee that the received signals fall in the detection region away from the decision boundaries. Here, the constellation points are not surrounded by the decision boundaries and therefore have a larger detection region that extend infinitely. The constraints are 
\begin{align*}
C1: & Re(y_i) \gtreqless {\sqrt{\Gamma_i^{DL}}}\sigma_iRe(d_i) 
\end{align*}
\begin{align*}
C2: & Im(y_i) \gtreqless {\sqrt{\Gamma_i^{DL}}}\sigma_iIm(d_i) 
\end{align*}
\end{itemize}

\subsubsection{Total Downlink Transmit PM Problem} 
By adopting the required constraints C1 and C2 for the corresponding group constellation points, the total  downlink transmit PM optimisation problem is expressed as
\begin{equation} 
\begin{split}
\mathcal{P}7: \, \underset{{\textbf{w}_k},P_j}{\text{min}} \quad & {\left\| \sum_{k=1}^{K} {\textbf{w}_k}d_k\right\|}^2 \\
\text{s.t.} \quad & \textrm{Constraints C1 and C2,} \forall i, \\
& C3: \frac {{P_j \mid{\textbf{f}_j^H}{\textbf{u}_j}\mid}^2} {I^{QAM}_j + \sigma_N ^2 {\lVert {\textbf{u}_j} \rVert}^2} \geq \Gamma_j^{UL}, \forall j. 
\end{split}
\end{equation}  
where \( I^{QAM}_j = \sum_{n\ne j}^{J} P_n \mid{\textbf{f}_n^H}{\textbf{u}_j}\mid ^2 + \sum_{k= 1}^{K} \mid{\textbf{u}_j^H}{\textbf{G}}{\textbf{w}_k{d_k}}\mid ^2 \).

\subsubsection{Total Uplink Transmit PM Problem} 
Similarly, the uplink PM problem can be written for the case of QAM as
\begin{equation} 
\begin{split}
\mathcal{P}8: \, \underset{{\textbf{w}_i},P_j}{\text{min}} \quad  & \sum_{j=1}^{J} {P_j} \\
\text{s.t.} \quad & \textrm{Constraints C1 and C2,} \forall i, \\
& C3: \frac {{P_j \mid{\textbf{f}_j^H}{\textbf{u}_j}\mid}^2} {I^{QAM}_j + \sigma_N ^2 {\lVert {\textbf{u}_j} \rVert}^2} \geq \Gamma_j^{UL}, \forall j. 
\end{split}
\end{equation}    		

\subsubsection{Multi-objective PM Problem}
Finally, we can design the MOOP for the case of QAM by employing the above constraints C1 and C2 as
\begin{equation} 
\begin{split}
\mathcal{P}9: \, \underset{{\textbf{w}_i},P_j,t}{\text{min}} \quad  & t \\
\text{s.t.} \quad & \textrm{Constraints C1 and C2,} \forall i, \\
& C3: \frac {{P_j \mid{\textbf{f}_j^H}{\textbf{u}_j}\mid}^2} {I^{QAM}_j + \sigma_N ^2 {\lVert {\textbf{u}_j} \rVert}^2} \geq \Gamma_j^{UL}, \forall j, \\
& C4: {\lambda_{a}}\left(R_a^* - R_a({\textbf{w}_i},P_j)\right) \leq t, \forall a\in\left\{1,2\right\}.
\end{split}
\end{equation}  

Again, it can be observed that unlike their conventional counterparts, all three optimizations above are convex and can be optimally solved using standard convex softwares like CVX \cite{grant2008cvx}.

\section{Multi-objective Optimization Problem with Imperfect CSI}
\subsection{Conventional Robust MOOP}
In this section we study the robustness of the system when the downlink, the uplink and the SI CSI is not perfectly known. For each channel, the actual CSI is modeled as
\begin{align} 
 {\textbf{h}}_i = {\check{\textbf{h}}}_i + {\Delta{\textbf{h}}}_i, \forall i, 
\end{align} 
\begin{align} 
 {\textbf{f}}_j = {\check{\textbf{f}}}_j + {\Delta{\textbf{f}}}_j, \forall j, 
\end{align} 
and,
\begin{align} 
 {\textbf{G}} = {\check{\textbf{G}}} + {\Delta{\textbf{G}}}.
\end{align} 

where \( {\check{\textbf{h}}}_i, {\check{\textbf{f}}}_j \) and \( {\check{\textbf{G}}} \) denote the downlink, the uplink and the SI CSI estimates known to the FD BS, respectively. And \( {\Delta{\textbf{h}}}_i, \forall i  {\Delta{\textbf{f}}}_j, \forall j \) and \( {\Delta{\textbf{G}}} \) represent the downlink, the uplink and the SI CSI uncertainties, respectively, which are assumed to be bounded such that
\begin{align}
  \left\|{\Delta{\textbf{h}}}_i \right\|^2 \leq \epsilon_{h,i}^2, \, \textrm{for some} \, \epsilon_{h,i} \geq 0,    \\
  \left\|{\Delta{\textbf{f}}}_j \right\|^2 \leq \epsilon_{f,j}^2,\, \textrm{for some} \, \epsilon_{f,i} \geq 0,  \\
  \left\|{\Delta{\textbf{G}}} \right\|^2 \leq \epsilon_G^2,\, \textrm{for some} \, \epsilon_{G} \geq 0.
\end{align}


We assume that the FD BS has no knowledge of \( {\Delta{\textbf{h}}}_i, {\Delta{\textbf{f}}}_j \) and \( {\Delta{\textbf{G}}} \) except for the error bounds, hence, we take the worst-case approach for our problem design. \par 
Henceforth, we focus on the multi-objective problem formulation since it is a generalisation of both the downlink and uplink optimisation problems. Therefore, the multi-object formulation of problem \( \mathcal{P}3 \) for imperfect CSI is
\begin{equation} 
\begin{split}
\mathcal{P}10: \quad \underset{{\textbf{w}_i},P_j,t}{\text{min}} \quad & t \\
\text{s.t.} \quad & \frac {{\left| \left(\check{\textbf{h}}_i + {\Delta{\textbf{h}}}_i\right)^H{\textbf{w}_i}\right|}^2 }{\sum_{k\ne i}^{K} {\left|\left(\check{\textbf{h}}_i + {\Delta{\textbf{h}}}_i\right)^H{\textbf{w}_k}\right|}^2  + \sigma_i ^2 }  \geq \Gamma_i^{DL}, \\
& \forall \left\|{\Delta{\textbf{h}}}_i \right\|^2 \leq \epsilon_{h,i}^2,\forall i, \\
& \frac {{P_j \left| \left(\check{\textbf{f}}_j + {\Delta{\textbf{f}}}_j\right)^H{\textbf{u}_j}\right|}^2} {\sum_{n\ne j}^{J} P_n\left| \left(\check{\textbf{f}}_j + {\Delta{\textbf{f}}}_j\right)^H{\textbf{u}_j}\right|^2  + C_j} \geq \Gamma_j^{UL},\\
& \forall \left\|{\Delta{\textbf{G}}} \right\|^2 \leq \epsilon_{G}^2,\forall \left\|{\Delta{\textbf{f}}}_j \right\|^2 \leq \epsilon_{f,j}^2,\forall j, \\
& {\lambda_{a}}\left(R_a^* - R_a\right) \leq t, \forall a\in\left\{1,2\right\}.
\end{split}
\end{equation}

where \( C_j =\sum_{k= 1}^{K}\left|{\textbf{u}_j^H}\left(\check{\textbf{G}} + {\Delta{\textbf{G}}}\right)\textbf{w}_k\right| ^2 + \sigma_N ^2 {\lVert {\textbf{u}_j} \rVert}^2 \)
\par
In the downlink and uplink SINR constraints, there are infinitely many inequalities which make the worst-case design particularly challenging. To make \( \mathcal{P}10 \) more tractable, we formulate the problem as a semidefinite program (SDP) then transform the constraints into linear matrix inequalities (LMI), which can be efficiently solved by existing solvers like CVX \cite{grant2008cvx}. The SDP transformation of problem \( \mathcal{P}10 \) is
\begin{equation} 
\begin{split}
& \underset{{\textbf{W}_i},P_j,t}{\text{min}} \quad t \\
\text{s.t.} \quad & \widetilde{\textrm{D1}}: \frac {\left(\check{\textbf{h}}_i + {\Delta{\textbf{h}}}_i\right)^H{\textbf{W}_i}\left(\check{\textbf{h}}_i + {\Delta{\textbf{h}}}_i\right)}{\sum_{k\ne i}^{K} \left((\check{\textbf{h}}_i + {\Delta{\textbf{h}}}_i)^H{\textbf{W}_k}(\check{\textbf{h}}_i + {\Delta{\textbf{h}}}_i) \right) + \sigma_i ^2 }  \geq \Gamma_i^{DL}, \\
& \forall \left\|{\Delta{\textbf{h}}}_i \right\|^2 \leq \epsilon_{h,i}^2,\forall i, \\
& \widetilde{\textrm{D2}}:  \frac {P_j \left(\check{\textbf{f}}_j + {\Delta{\textbf{f}}}_j\right)^H{\textbf{U}_j}\left(\check{\textbf{f}}_j + {\Delta{\textbf{f}}}_j\right)} {\sum_{n\ne j}^{J} P_n \left(\check{\textbf{f}}_j + {\Delta{\textbf{f}}}_j\right)^H{\textbf{U}_j}\left(\check{\textbf{f}}_j + {\Delta{\textbf{f}}}_j\right) + \widetilde{C}_j } \geq \Gamma_j^{UL}, \\
& \forall \left\|{\Delta{\textbf{G}}} \right\|^2 \leq \epsilon_{G}^2,\forall \left\|{\Delta{\textbf{f}}}_j \right\|^2 \leq \epsilon_{f,j}^2,\forall j, \\
& \widetilde{\textrm{D3}}: {\lambda_{a}}\left(R_a^* - R_a\right) \leq t, \forall a\in\left\{1,2\right\}. \\
& \widetilde{\textrm{D4}}: {\textbf{W}_i} \succeq 0, \forall i.
\end{split}
\end{equation}

where, \\
\( \widetilde{C}_j = \textrm{Tr}\left\{\left(\check{\textbf{G}} + {\Delta{\textbf{G}}}\right)\sum_{k= 1}^{K}{\textbf{W}_k}\left(\check{\textbf{G}} + {\Delta{\textbf{G}}}\right)^H{\textbf{U}}_j\right\} +  \sigma_N ^2  \textrm{Tr}\left\{{\textbf{U}}_j \right\} \) and we define \( \textbf{W}_i = \textbf{w}_i\textbf{w}_i^H \) and \( \textbf{U}_j = \textbf{u}_j\textbf{u}_j^H  \). 
\par By applying the S-procedure as in \cite{boyd2004convex} we can convert these constraints into LMIs. First, we can rearrange constraint \( \widetilde{\textrm{D1}} \) into
\begin{align}
  \left(\check{\textbf{h}}_i + {\Delta{\textbf{h}}}_i\right)^H{\textbf{Q}_i}\left(\check{\textbf{h}}_i + {\Delta{\textbf{h}}}_i\right) - \Gamma_i^{DL} \sigma_i ^2   \geq 0, \forall i,
\end{align}
where, we introduce 
\begin{align*}
{\textbf{Q}_i} \triangleq {\textbf{W}_i} - \Gamma_i^{DL} \sum_{k\ne i}^{K}{\textbf{W}_k}, \forall i 
\end{align*}
and then for constraint \( \widetilde{\textrm{D2}} \), let's define two vectors \( \widetilde{\textbf{f}} \)  and \( \widetilde{\Delta{\textbf{f}}} \) as
\begin{align}
\widetilde{\textbf{f}} = {\begin{bmatrix} \check{\textbf{f}}_j \\ \vdots \\ \check{\textbf{f}}_J \end{bmatrix}} \in \mathbb{C}^{NJ \times 1}, \widetilde{\Delta{\textbf{f}}} = {\begin{bmatrix} {\Delta{\textbf{f}}}_j \\ \vdots \\ {\Delta{\textbf{f}}}_J \end{bmatrix}} \in \mathbb{C}^{NJ \times 1}
\end{align}  
hence, we can define any \( \check{\textbf{f}}_j  = \textbf{B}_j\widetilde{\textbf{f}} \) and \( {\Delta{\textbf{f}}}_j = \textbf{B}_j\widetilde{\Delta{\textbf{f}}},\) for \( j = 1,\hdots,J, \) with \( \textbf{B}_j \in \mathbb{R}^{ N \times NJ} \) defined as \( \textbf{B}_j = {\begin{bmatrix} \textbf{B}_{j,1},\hdots,\textbf{B}_{j,J} \end{bmatrix}}, \) where \( \textbf{B}_{j,j} = \textbf{I}_N \) and \( \textbf{B}_{j,n} = \textbf{0}_N,\) for \( n = 1,\hdots,J, n\ne j. \) We have \( \textbf{I}_N \) and \( \textbf{0}_N \) to be an \( N \times N \) identity matrix and zero matrix, respectively. \par 
Now constraint \( \widetilde{\textrm{D2}} \) can be rewritten as
\begin{align}
\frac {P_j \left((\textbf{B}_j\widetilde{\textbf{f}} + \textbf{B}_j\widetilde{\Delta{\textbf{f}}} )^H{\textbf{U}_j}(\textbf{B}_j\widetilde{\textbf{f}} + \textbf{B}_j\widetilde{\Delta{\textbf{f}}})\right)} {\sum_{n\ne j}^{J} P_n \left((\textbf{B}_n\widetilde{\textbf{f}} + \textbf{B}_n\widetilde{\Delta{\textbf{f}}} )^H{\textbf{U}_j}(\textbf{B}_n\widetilde{\textbf{f}} + \textbf{B}_n\widetilde{\Delta{\textbf{f}}})\right) + \widetilde{C}_j } \geq \Gamma_j^{UL}, \forall j 
\end{align}      
and can be simplified to give
\begin{align}
\frac{\left(\widetilde{\textbf{f}} + \widetilde{\Delta{\textbf{f}}} \right)^H{\textbf{Z}_j}\left(\widetilde{\textbf{f}} +\widetilde{\Delta{\textbf{f}}}\right)}{\textrm{Tr}\left\{\left(\check{\textbf{G}} + {\Delta{\textbf{G}}}\right)\sum_{k= 1}^{K}{\textbf{W}_k}\left(\check{\textbf{G}} + {\Delta{\textbf{G}}}\right)^H{\textbf{U}}_j\right\} + \sigma_N ^2  \textrm{Tr}\left\{{\textbf{U}}_j \right\}}  \geq \Gamma_j^{UL}
\end{align}  
where we introduce 
\begin{align*}
{\textbf{Z}_j} \triangleq P_j\textbf{B}_j^T{\textbf{U}_j}\textbf{B}_j - \Gamma_j^{UL}\sum_{n\ne j}^{J} P_n\textbf{B}_n^T{\textbf{U}_j}\textbf{B}_n, \forall j
\end{align*}
we further simplify (35) by introducing slack variables \( s_j > 0, \forall{j} \) \cite{boyd2004convex}, such that (35) can be written as the following two constraints
\begin{align}
    \left(\widetilde{\textbf{f}} + \widetilde{\Delta{\textbf{f}}} \right)^H{\textbf{Z}_j}\left(\widetilde{\textbf{f}} +\widetilde{\Delta{\textbf{f}}}\right) \geq s_j\Gamma_j^{UL}, \forall{j}, 
\end{align}
\begin{align}
    \textrm{Tr}\left\{\left(\check{\textbf{G}} + {\Delta{\textbf{G}}}\right)\sum_{k= 1}^{K}{\textbf{W}_k}\left(\check{\textbf{G}} + {\Delta{\textbf{G}}}\right)^H{\textbf{U}}_j\right\} + \sigma_N ^2  \textrm{Tr}\left\{{\textbf{U}}_j \right\} \leq s_j, \forall j.
\end{align}
    
Next, we review the definitions of the S-procedure.

\textbf{Lemma 1. (S-procedure \cite{boyd2004convex}):} Let \( g_l(\textbf{x}), \) \(l =1,2, \) be defined as
\begin{align*}
g_l(\textbf{x}) = \textbf{x}^H \textbf{A}_l\textbf{x} + 2Re\left\{\textbf{b}_l^H\textbf{x}\right\} + c_l
\end{align*}  
where \( \textbf{A}_l \in \mathbb{C}^{n\times n}, \textbf{b}_l \in \mathbb{C}^n \) and \( c_l \in \mathbb{R} \). Then, the implication of \( g_1(\textbf{x}) \geq 0 \Rightarrow g_2(\textbf{x}) \geq 0 \) holds if and only if there exists a \( \lambda \geq 0 \) such that
\begin{align*}
\delta {\begin{bmatrix} \textbf{A}_1 & \textbf{b}_1 \\ \textbf{b}_1^H & c_1 \end{bmatrix}} - {\begin{bmatrix} \textbf{A}_2 & \textbf{b}_2 \\ \textbf{b}_2^H & c_2 \end{bmatrix}} \succeq 0
\end{align*}  
provided there exists a point \( \hat{\textbf{x}} \) with \( g_1(\hat{\textbf{x}}) > 0. \) \\
\par Following Lemma 1 and using the fact that \( \textrm{Tr}\left\{\textbf{ABCD}  \right\} = \textrm{vec}\left(\textbf{A}^H \right)^H\left(\textbf{D}^H \otimes \textbf{B} \right)\textrm{vec}\left(\textbf{C} \right)\), constraints (32), (36) and (37) can be expanded as
\begin{align}
{\Delta{\textbf{h}}}_i^H{\textbf{Q}_i}{\Delta{\textbf{h}}}_i + 2Re\left\{ \check{\textbf{h}}_i^H {\textbf{Q}_i} {\Delta{\textbf{h}}}_i \right\} + \check{\textbf{h}}_i^H {\textbf{Q}_i}\check{\textbf{h}}_i - \Gamma_i^{DL} \sigma_i ^2   \geq 0, \forall i 
\end{align} 
\begin{align} 
\widetilde{\Delta{\textbf{f}}}^H{\textbf{Z}_j}\widetilde{\Delta{\textbf{f}}} + 2Re\left\{ \widetilde{\textbf{f}}^H {\textbf{Z}_j} \widetilde{\Delta{\textbf{f}}} \right\} + \widetilde{\textbf{f}}^H {\textbf{Z}_j}\widetilde{\textbf{f}} - s_j\Gamma_j^{UL} \geq 0, \forall j,
\end{align} 
\begin{multline}
{\Delta{\textbf{g}}}^H \left({\textbf{U}}_j \otimes\sum_{k= 1}^{K}{\textbf{W}_k}\right){\Delta{\textbf{g}}}+ 2Re\left\{ \check{\textbf{g}}^H \left({\textbf{U}}_j \otimes\sum_{k= 1}^{K}{\textbf{W}_k}\right){\Delta{\textbf{g}}} \right\} \\ + \check{\textbf{g}}^H \left({\textbf{U}}_j \otimes\sum_{k= 1}^{K}{\textbf{W}_k}\right)\check{\textbf{g}} + \sigma_N ^2  \textrm{Tr}\left\{{\textbf{U}}_j \right\} - s_j \leq 0, \forall j
\end{multline} 

\begin{figure*}

\begin{equation} 
\begin{split}
{\mathcal{P}11}: \quad  \underset{{\textbf{W}_i},P_j,t}{\text{min}} \quad & t \\
\text{s.t.}  \quad & {\begin{bmatrix} {\delta_i{\textbf{I}} + \textbf{Q}_i} & {\textbf{Q}_i}\check{\textbf{h}}_i \\ \check{\textbf{h}}_i^H{\textbf{Q}_i} &  \check{\textbf{h}}_i^H {\textbf{Q}_i}\check{\textbf{h}}_i  - \Gamma_i^{DL} \sigma_i ^2 - \delta_i \epsilon_{h,i}^2 \end{bmatrix}} \succeq 0, \forall i, \\
& {\begin{bmatrix} {\mu_j{\textbf{I}} + \textbf{Z}_j} & {\textbf{Z}_j}\widetilde{\textbf{f}} \\ \widetilde{\textbf{f}}^H{\textbf{Z}_j} &  \widetilde{\textbf{f}}^H {\textbf{Z}_j}\widetilde{\textbf{f}}  - s_j\Gamma_j^{UL} - \mu_j \epsilon_{f,j}^2 \end{bmatrix}} \succeq 0, \forall j, \\
& {\begin{bmatrix} {\rho{\textbf{I}} - \left({\textbf{U}}_j \otimes\sum_{k= 1}^{K}{\textbf{W}_k}\right)} & - \left({\textbf{U}}_j \otimes\sum_{k= 1}^{K}{\textbf{W}_k}\right)\check{\textbf{g}} \\ -\check{\textbf{g}}^H \left({\textbf{U}}_j \otimes\sum_{k= 1}^{K}{\textbf{W}_k}\right) & s_j - \check{\textbf{g}}^H \left({\textbf{U}}_j \otimes\sum_{k= 1}^{K}{\textbf{W}_k}\right)\check{\textbf{g}} - \sigma_N ^2  \textrm{Tr}\left\{{\textbf{U}}_j \right\} - \rho \epsilon_{G}^2 \end{bmatrix}} \succeq 0, \forall j, \\
& {\lambda_{a}}\left(R_a^* - R_a\right) \leq t, \forall a\in\left\{1,2\right\}, \\
& {\textbf{W}_i} \succeq 0,\delta_i \geq 0,\mu_j \geq 0, \rho \geq 0, s_j > 0,  \forall i,j.
\end{split}
\end{equation}
\hrulefill
\end{figure*}

we define \(  {\check{\textbf{g}}} = \textrm{vec}\left({\check{\textbf{G}}}^H\right) \) and \( {\Delta{\textbf{g}}} = \textrm{vec}\left({\Delta{\textbf{G}}}^H\right) \) where, \( \textrm{vec}\left( \cdot \right)\)  stacks the columns of a matrix into a vector and \( \otimes \) stands for Kronecker product. 

\par Hence, by exploiting the S-procedure in Lemma 1, (38), (39) and (40) can be formulated as LMIs and the conventional robust optimisation problem \( \mathcal{P}10 \) can be reformulated as shown in (41).
\par The problem \( {\mathcal{P}11} \) is convex, and can be efficiently solved using CVX \cite{grant2008cvx}. The resulting optimal values obtained from \( {\mathcal{P}11} \) provide a lower bound for the conventional power minimisation problem.
\par Note that the problem \( {\mathcal{P}11} \) is a relaxed form of \( {\mathcal{P}10} \). When the relaxation in \({\mathcal{P}11}\) is tight, i.e. \({\mathcal{P}11}\) returns all rank-one solutions \(( {\textbf{W}_i}) \), then the optimal solution \( ({\textbf{w}_i}) \) to solve \({\mathcal{P}10}\) can be obtained by matrix decomposition or randomisation as in \cite{luo2010semidefinite}, such that \( {\textbf{W}_i} = \textbf{w}_i\textbf{w}_i^H, \forall i. \) Otherwise, the required power in original problem \( {\mathcal{P}10} \) is always higher than that in \( {\mathcal{P}11} \).

\subsection{Robust MOOP based on Constructive Interference}
To study the robustness of the proposed system for the case of constructive interference, we first formulate \( \mathcal{P}6 \) as a virtual multicast problem \cite{sidiropoulos2006transmit}. To facilitate this, we simply incorporate each user's channel with its respective data symbol i.e. \( {\widetilde{\textbf{h}}_i} = {\textbf{h}_i}{e^{j({\phi}_1 - {\phi}_i)}} \) and let \( \textbf{w}=\sum_{k= 1}^{K} {\textbf{w}_k}{e^{j({\phi}_k - {\phi}_1)}} \). Following this the multicast formulation of problem \( \mathcal{P}6 \) can be written as 
\begin{equation} 
\begin{split}
{\mathcal{P}12}: \quad & \underset{\textbf{w},P_j,t}{\text{min}} \quad t \\
\text{s.t.} \quad & \left| Im \left({{\widetilde{\textbf{h}}_i}^H} \textbf{w}\right) \right| \leq \left(Re \left({{\widetilde{\textbf{h}}_i}^H} \textbf{w}\right) - {\sqrt{\Gamma_i^{DL} \sigma_i^2} }\right) \tan \theta, \forall i, \\
& \frac {{P_j \left|{\textbf{f}_j^H}{\textbf{u}_j}\right|}^2} {\sum_{n\ne j}^{J} P_n \left|{\textbf{f}_n^H}{\textbf{u}_j}\right| ^2 + \left|{\textbf{u}_j^H}{\textbf{G}}\textbf{w}\right| ^2 + \sigma_N ^2 {\lVert {\textbf{u}_j} \rVert}^2} \geq \Gamma_j^{UL}, \forall j, \\
& {\lambda_{a}}\left(R_a^* - R_a\right) \leq t, \forall a\in\left\{1,2\right\}.
\end{split}
\end{equation}
\par Based on the multicast formulation \( {\mathcal{P}12} \), for the worst-case design we model the imperfect CSI as 
\begin{align} 
\widetilde{\textbf{h}}_i = {\widetilde{\textbf{h}}}_i + \Delta\widetilde{\textbf{h}}_i, \forall i 
\end{align}  
where \( {\widetilde{\textbf{h}}}_i \) denotes the downlink CSI estimate known to the FD BS. And \( {\Delta\widetilde{\textbf{h}}}_i \) is the downlink CSI uncertainty which is bounded such that \(  \left\|{\Delta\widetilde{\textbf{h}}}_i \right\|^2 \leq \epsilon_{h,i}^2 \). And we model the uplink and the SI CSI as in Section V-A, respectively. The robust formulation of problem \( {\mathcal{P}12} \) is 
\begin{equation} 
\begin{split}
{\mathcal{P}13}:\,\underset{\textbf{w},P_j,t}{\text{min}} \quad & t \\
\text{s.t.} \quad & \left| Im \left(({\widetilde{\textbf{h}}}_i + \Delta\widetilde{\textbf{h}}_i)^H \textbf{w}\right) \right| \\
& \leq \left(Re \left(({\widetilde{\textbf{h}}}_i + \Delta\widetilde{\textbf{h}}_i)^H \textbf{w}\right) - {\sqrt{\Gamma_i^{DL} \sigma_i^2} }\right) \tan \theta, \\
& \forall \left\|{\Delta{\textbf{h}}}_i \right\|^2 \leq \epsilon_{h,i}^2,\forall i, \\    
& \frac {{P_j \left| \left(\check{\textbf{f}}_j + {\Delta{\textbf{f}}}_j\right)^H{\textbf{u}_j}\right|}^2} {\sum_{n\ne j}^{J} P_n\left| \left(\check{\textbf{f}}_j + {\Delta{\textbf{f}}}_j\right)^H{\textbf{u}_j}\right|^2  + I_j} \geq \Gamma_j^{UL}, \\
& \forall \left\|{\Delta{\textbf{G}}} \right\|^2 \leq \epsilon_{G}^2,\forall \left\|{\Delta{\textbf{f}}}_j \right\|^2 \leq \epsilon_{f,j}^2,\forall j, \\
& {\lambda_{a}}\left(R_a^* - R_a\right) \leq t, \forall a\in\left\{1,2\right\}.
\end{split}
\end{equation}
where \( I_j = \left|{\textbf{u}_j^H}\left(\check{\textbf{G}} + {\Delta{\textbf{G}}}\right)\textbf{w}\right| ^2 + \sigma_N ^2 {\lVert {\textbf{u}_j} \rVert}^2 \).

First, let's consider the downlink SINR constraint. For convenience we separate the real and imaginary part of the complex notations and represent them as real valued numbers. Let
\begin{align}
{\underline{\textbf{w}}} & \triangleq {\begin{bmatrix} Re({\textbf{w}}) \\ Im({\textbf{w}}) \end{bmatrix}},\\
{\underline{\widetilde{\textbf{h}}}}_i & \triangleq {\begin{bmatrix} Im({\widetilde{\textbf{h}}_i})^H & Re({\widetilde{\textbf{h}}_i})^H \end{bmatrix}}, \\
{\Delta\underline{\widetilde{\textbf{h}}}}_i & \triangleq {\begin{bmatrix} Im(\Delta{\widetilde{\textbf{h}}_i})^H & Re(\Delta{\widetilde{\textbf{h}}_i})^H \end{bmatrix}}, \\
{\boldsymbol{\Pi}} & \triangleq {\begin{bmatrix} {\textbf{0}}_N & -{\textbf{I}}_N \\ {\textbf{I}}_N & {\textbf{0}}_N\end{bmatrix}}. 
\end{align}
\par Where, \( {\textbf{0}}_N \) and \( {\textbf{I}}_N \) denote \( N \) x \( N \)  all-zero matrix and identity matrix, respectively. \par With the new notations we can express the real and imaginary terms of downlink SINR constraint in \( {\mathcal{P}13} \) as:
\begin{align}
\textrm{Im} ({\widetilde{\textbf{h}}_i^H} {\textbf{w}}) = ({\underline{\widetilde{\textbf{h}}}}_i + {\Delta\underline{\widetilde{\textbf{h}}}}_i){\underline{\textbf{w}}}, && \textrm{Re} ({\widetilde{\textbf{h}}_i^H} {\textbf{w}}) = ({\underline{\widetilde{\textbf{h}}}}_i + {\Delta\underline{\widetilde{\textbf{h}}}}_i){\boldsymbol{\Pi}}{\underline{\textbf{w}}}
\end{align}
\par From the definition of the error bound, we have \( \left\|{\Delta\widetilde{\textbf{h}}}_i \right\|^2 \leq \epsilon_{h,i}^2 \), the downlink SINR constraint can be guaranteed by the following constraint
\begin{multline}
\underset{\lVert{\Delta\widetilde{\textbf{h}}}_i \rVert^2 \leq \epsilon_{h,i}^2}{\text{max}} \quad \left|\left({\underline{\widetilde{\textbf{h}}}}_i + {\Delta\underline{\widetilde{\textbf{h}}}}_i\right){\underline{\textbf{w}}} \right| \\
- \left(\left({\underline{\widetilde{\textbf{h}}}}_i + {\Delta\underline{\widetilde{\textbf{h}}}}_i\right){\boldsymbol{\Pi}}{\underline{\textbf{w}}} - {\sqrt{\Gamma_i^{DL} \sigma_i^2} }\right) \tan \theta \leq 0, \forall i
\end{multline}

\par  Hence, by considering the absolute value term, (50) is equivalent to the following two constraints
\begin{multline}
\underset{\lVert{\Delta\widetilde{\textbf{h}}}_i \rVert^2 \leq \epsilon_{h,i}^2}{\text{max}} \quad {\underline{\widetilde{\textbf{h}}}}_i{\underline{\textbf{w}}} + {\Delta\underline{\widetilde{\textbf{h}}}}_i{\underline{\textbf{w}}} - \left({\underline{\widetilde{\textbf{h}}}}_i + {\Delta\underline{\widetilde{\textbf{h}}}}_i\right){\boldsymbol{\Pi}}{\underline{\textbf{w}}}\tan \theta \\ + {\sqrt{\Gamma_i^{DL} \sigma_i^2} } \tan \theta \leq 0, \forall i
\end{multline}

\begin{multline}
\underset{\lVert{\Delta\widetilde{\textbf{h}}}_i \rVert^2 \leq \epsilon_{h,i}^2}{\text{max}} \quad - {\underline{\widetilde{\textbf{h}}}}_i{\underline{\textbf{w}}} - {\Delta\underline{\widetilde{\textbf{h}}}}_i{\underline{\textbf{w}}} - \left({\underline{\widetilde{\textbf{h}}}}_i + {\Delta\underline{\widetilde{\textbf{h}}}}_i\right){\boldsymbol{\Pi}}{\underline{\textbf{w}}}\tan \theta \\ + {\sqrt{\Gamma_i^{DL} \sigma_i^2} } \tan \theta \leq 0, \forall i
\end{multline}
Therefore, the the robust formulations of (51) and (52) are given by 
\begin{multline}
{\underline{\widetilde{\textbf{h}}}}_i{\underline{\textbf{w}}} - {\underline{\widetilde{\textbf{h}}}}_i{\boldsymbol{\Pi}}{\underline{\textbf{w}}}\tan \theta + \epsilon_{h,i}\left\| {\underline{\textbf{w}}} - {\boldsymbol{\Pi}}{\underline{\textbf{w}}}\tan \theta \right\| \\
+ {\sqrt{\Gamma_i^{DL} \sigma_i^2} } \tan \theta \leq 0, \forall i
\end{multline}
\begin{multline}
-{\underline{\widetilde{\textbf{h}}}}_i{\underline{\textbf{w}}} - {\underline{\widetilde{\textbf{h}}}}_i{\boldsymbol{\Pi}}{\underline{\textbf{w}}}\tan \theta + \epsilon_{h,i}\left\| {-\underline{\textbf{w}}} - {\boldsymbol{\Pi}}{\underline{\textbf{w}}}\tan \theta \right\| \\
+ {\sqrt{\Gamma_i^{DL} \sigma_i^2} } \tan \theta \leq 0, \forall i
\end{multline}

\par Next, we consider the uplink SINR constraint in problem (44). Following equations (33) and (34) in Section V-A, the uplink SINR constraint can be rewritten as
\begin{multline}
\frac{\left(\widetilde{\textbf{f}} + \widetilde{\Delta{\textbf{f}}} \right)^H{\textbf{Z}_j}\left(\widetilde{\textbf{f}} +\widetilde{\Delta{\textbf{f}}}\right)}{\left|{\textbf{u}_j^H}\check{\textbf{G}}\textbf{w} + {\textbf{u}_j^H}{\Delta{\textbf{G}}}\textbf{w} \right| ^2 + \sigma_N ^2 {\lVert {\textbf{u}_j} \rVert}^2 }  \geq \Gamma_j^{UL}, \\ \forall \left\|{\Delta{\textbf{G}}} \right\|^2 \leq \epsilon_{G}^2,\forall \left\|{\Delta{\textbf{f}}}_j \right\|^2 \leq \epsilon_{f,j}^2,\forall j.
\end{multline}
and we note that (55) can be guaranteed by the following constraints

\begin{multline}
 \underset{\lVert{\Delta\widetilde{\textbf{f}}}_j \rVert^2 \leq \epsilon_{f,j}^2}{\text{max}} \quad\left(\widetilde{\textbf{f}} + \widetilde{\Delta{\textbf{f}}} \right)^H{\textbf{Z}_j}\left(\widetilde{\textbf{f}} +\widetilde{\Delta{\textbf{f}}}\right) - \Gamma_j^{UL} \left({c}_j + \sigma_N ^2 {\lVert {\textbf{u}_j} \rVert}^2 \right) \\ 
 \geq 0, \forall j
\end{multline}
\begin{align}
    \underset{\lVert{\Delta\widetilde{\textbf{G}}} \rVert^2 \leq \epsilon_{G}^2}{\text{max}} \quad \left|{\textbf{u}_j^H}\check{\textbf{G}}\textbf{w} + {\textbf{u}_j^H}{\Delta{\textbf{G}}}\textbf{w} \right| ^2 \leq c_j, \forall j
\end{align}
where \( c_j > 0, \forall{j} \) are introduced as slack variables \cite{boyd2004convex}. 
\par Similar procedure as in Section V-A can be applied to (56). By exploiting the S-procedure in Lemma 1, (56) can be expanded and converted into a LMI as shown below
\begin{multline}
{\begin{bmatrix} {\mu_j{\textbf{I}_N} + \textbf{Z}_j} & {\textbf{Z}_j}\widetilde{\textbf{f}} \\ \widetilde{\textbf{f}}^H{\textbf{Z}_j} &  \widetilde{\textbf{f}}^H {\textbf{Z}_j}\widetilde{\textbf{f}}  - \Gamma_j^{UL}c_j - \Gamma_j^{UL}\sigma_N ^2 \textrm{Tr}({\textbf{U}_j}) - \mu_j \epsilon_{f,j}^2 \end{bmatrix}} \\ \succeq 0, \forall j,
\end{multline}

We note that by using the fact that \( \left\| x + y \right\|^2 \leq  \left(\left\| x \right\|+ \left\| y \right\| \right)^2 \), (57) can always be guaranteed by the following constraint
\begin{align}
    \underset{\lVert{\Delta\widetilde{\textbf{G}}} \rVert^2 \leq \epsilon_{G}^2}{\text{max}} \quad \left(\left|{\textbf{u}_j^H}\check{\textbf{G}}\textbf{w} \right| + \left| {\textbf{u}_j^H}{\Delta{\textbf{G}}}\textbf{w} \right| \right)^2 \leq c_j, \forall{j}
\end{align}
whose robust formulation is given by  
\begin{align}
    \left(\left|{\textbf{u}_j^H}\check{\textbf{G}}\textbf{w} \right| + \epsilon_{G}\left| {\textbf{u}_j^H}\textbf{w} \right| \right)^2 \leq c_j, \forall{j}
\end{align}
Futhermore, we define \( {\underline{\textbf{Y}}}_j  \triangleq {\begin{bmatrix}  Re({\textbf{u}_j^H}{\textbf{G}}) & -Im({\textbf{u}_j^H}{\textbf{G}}) \\ Im({\textbf{u}_j^H}{\textbf{G}}) & Re({\textbf{u}_j^H}{\textbf{G}})  \end{bmatrix}}\) and \( {\underline{\textbf{U}}}_j \triangleq {\begin{bmatrix}  Re({\textbf{u}_j^H}) & -Im({\textbf{u}_j^H}) \\ Im({\textbf{u}_j^H}) & Re({\textbf{u}_j^H})  \end{bmatrix}}  \), therefore, the  constraint (60) can be written in terms of real valued numbers as 

\begin{align}
    \left(\left|{\underline{\textbf{Y}}}_j{\underline{\textbf{w}}} \right| + \epsilon_{G}\left| {\underline{\textbf{U}}}_j{\underline{\textbf{w}}} \right| \right)^2 \leq c_j, \forall{j}
\end{align}
Therefore, the robust optimisation problem based on CI is
\begin{equation} 
\begin{split}
{\mathcal{P}14}: & \, \underset{{\underline{\textbf{w}}},P_j,t}  {\text{min}} \quad t \\
\text{s.t.}  \\
&{\underline{\widetilde{\textbf{h}}}}_i{\underline{\textbf{w}}} - {\underline{\widetilde{\textbf{h}}}}_i{\boldsymbol{\Pi}}{\underline{\textbf{w}}}\tan \theta + \epsilon_{h,i}\left\| {\underline{\textbf{w}}} - {\boldsymbol{\Pi}}{\underline{\textbf{w}}}\tan \theta \right\| \\
& \qquad \qquad \qquad \qquad \qquad \qquad \qquad \leq {\sqrt{\Gamma_i^{DL} \sigma_i^2} } \tan \theta, \forall i \\
& -{\underline{\widetilde{\textbf{h}}}}_i{\underline{\textbf{w}}} - {\underline{\widetilde{\textbf{h}}}}_i{\boldsymbol{\Pi}}{\underline{\textbf{w}}}\tan \theta + \epsilon_{h,i}\left\| {-\underline{\textbf{w}}} - {\boldsymbol{\Pi}}{\underline{\textbf{w}}}\tan \theta \right\| \\
& \qquad \qquad \qquad \qquad \qquad \qquad \qquad \leq {\sqrt{\Gamma_i^{DL} \sigma_i^2} } \tan \theta, \forall i, \\
& {\begin{bmatrix} {\mu_j{\textbf{I}_N} + \textbf{Z}_j} & {\textbf{Z}_j}\widetilde{\textbf{f}} \\ \widetilde{\textbf{f}}^H{\textbf{Z}_j} &  \widetilde{\textbf{f}}^H {\textbf{Z}_j}\widetilde{\textbf{f}}  - \Gamma_j^{UL}c_j - \Gamma_j^{UL}\sigma_N ^2 \textrm{Tr}({\textbf{U}_j}) - \mu_j \epsilon_{f,j}^2 \end{bmatrix}} \\
& \qquad \qquad \qquad \qquad \qquad \qquad \qquad \qquad \qquad \succeq 0, \forall j, \\    & \left(\left|{\underline{\textbf{Y}}}_j{\underline{\textbf{w}}} \right| + \epsilon_{G}\left| {\underline{\textbf{U}}}_j{\underline{\textbf{w}}} \right| \right)^2 \leq c_j, \forall{j}, \\
& {\lambda_{a}}\left(R_a^* - R_a\right) \leq t, \forall a\in\left\{1,2\right\}, \\
& \mu_j  \geq 0,\, c_j > 0, \forall j. 
\end{split}
\end{equation}

Note that problem \( {\mathcal{P}14} \) is a convex problem and thus can be optimally solved using standard convex softwares like CVX \cite{grant2008cvx}. After we obtain the optimal \({\underline{\textbf{w}}}^* \) and \( P_j^* \), the robust solution \({\textbf{w}}^* \) can be obtained from the relation in (45).

\section{Computational Complexity Analysis}
In this Section, we mathematically characterize the computational complexity of the conventional and proposed schemes based on MOOP formulations.

\subsection{Transmit Complexity}
We note that the convex MOOP formulations \( {\mathcal{P}3},{\mathcal{P}6}, {\mathcal{P}11} \) and \( {\mathcal{P}14} \) involve only LMI and second-order cone (SOC) constraints. As such, the problems can be sovled by a standard interior-point method (IPM) \cite{ben2001lectures}. Therefore we can use the worst-case runtime to analyse the complexity of the conventional and the proposed CI schemes. \par Following \cite{wang2014outage} and \cite{khandaker2017probabilistically}, the complexity of a generic IPM for solving problems like \({\mathcal{P}3},{\mathcal{P}6}, {\mathcal{P}11} \) and \( {\mathcal{P}14} \) involve the computation of a per-optimization cost. In each iteration, the computation cost is dominated by (i)
the formation of the coefficient matrix of the linear
system, and (ii) the factorization of the coefficient matrix. The cost of formation of the coefficient (\( C_{form }\)) matrix
is on the order of  
\begin{align*}
    C_{form} = \underset{\textrm{due to the LMI}}{\underbrace{n \sum_{a= 1}^{A} k_a^3 + n^2 \sum_{a= 1}^{A} k_a^2}} + \underset{\textrm{due to the SOC}}{\underbrace{n \sum_{a=A+1}^{B} k_a^2 }}  
\end{align*}
while the cost of factorizing (\( C_{fact }\))  is on the order of
\( C_{fact} = n^3\) (\(n=\) number of decision variables). Hence, the total computation cost per optimization is on the
order of \(C_{form} + C_{fact} \) \cite{wang2014outage}. We assume for the sake of simplicity that the decision variables in \({\mathcal{P}3},{\mathcal{P}6}, {\mathcal{P}11} \) and \( {\mathcal{P}14} \) are real-valued.
\par Hence, using these concepts, we now analyse the comutational complexity of \({\mathcal{P}3},{\mathcal{P}6}, {\mathcal{P}11} \) and \( {\mathcal{P}14} \). First we consider SDP formulation of \( {\mathcal{P}3} \), which has \(K\) LMI (trace) constraints of size 1, \(J\) LMI (trace) constraints of size 1, \(K\) SOC constraints of size \(N\), \(J\) LMI (trace) constraints of size 1 and \(K\) LMI (trace) constraints of size \(N\). Therefore, the complexity of the SDP formulation of \( {\mathcal{P}3} \) is on the order shown in the first row of Table I. Similarly, we can determine the complexity order of the formulations \({\mathcal{P}6}, {\mathcal{P}11} \) and \( {\mathcal{P}14} \) as shown in Table I, respectively. From Table I, we can show that the proposed MOOP formulation \( {\mathcal{P}6}\) has lower complexity than the SDP formulation of \( {\mathcal{P}3}\) since it has lower order of variables to compute i.e lower cost of factorization (\(C_{fact}\)). Also, we can straightforwardly show that for the robust MOOP, the proposed formulation \( {\mathcal{P}14}\) has a lower complexity than the conventional formulation \( {\mathcal{P}11}\) since \( {\mathcal{P}11}\) involves a more complicated set of constraints (5 LMI constraints and 1 SOC constrsint). This is also consistent with our simulation results in the following Section.

\begin{table}
\renewcommand{\arraystretch}{1.0}
\caption{Complexity Analysis of the MOOP Formulations}
\label{1}
\begin{tabular}{l||l}
\hline
\bfseries MOOP & \bfseries Complexity Order \\ 
\hline\hline
\({\mathcal{P}3}\)(SDP) &\( \mathcal{O}((KN^2+J)[ K(1+N^3) + 2J  + (KN^2+J) (K(1+N^2) \) \\ & \( + 2J) + KN^2 + (KN^2+J)^2]) \)\\
\hline \\
\({\mathcal{P}6}\) & \(\mathcal{O}( (KN+J)[2J(1+(KN+J)) + 2KN^2 + (KN+J)^2]) \)\\
\hline \hline \\
\({\mathcal{P}11}\) & \( \mathcal{O}((KN^2+J)[K(N+1)^2 + J(NJ+1)^3 + J(N^2+1)^3 \) \\ & \( + J + KN^3 + (KN^2+J)(K(N+1)^2 + J(NJ+1)^2 \) \\ & \( + J(N^2+1)^2 + J + KN^2) + (KN^2+J)(KN^2) \) \\ & \( + (KN^2+J)^2] ) \)\\
\hline \\
\({\mathcal{P}14}\) & \( \mathcal{O}((2N+J)[J(NJ+1)^3+J(N+1)^3+J \) \\ & \( + (2N+J)(J(NJ+1)^2+J+12N^2) + (2N+J)^2)  ]  ) \) \\
\hline
\end{tabular}
\end{table}

\par At this point, we emphasize that as the MOOP formulations in \({\mathcal{P}3} \) and \( {\mathcal{P}11} \) are data independent, they only need to be applied once during each channel coherence time. While as the proposed MOOP formulations in \({\mathcal{P}6} \) and \( {\mathcal{P}14} \) are data dependent, they need to be run on a symbol by symbol basis. In the following section we compare the resulting transmit complexity of conventional and proposed MOOP approaches for both slow and fast fading scenarios, and show that the average execution time per downlink frames is comparable for both techniques.

\subsection{Receiver Complexity}
At the receiver side, for the case of the conventional beamforming, the downlink users in our FD system scenario need to equalize the composite channel \( {\textbf{h}_i^H}{\textbf{w}_i}^* \) to recover their data symbols, where \(\left\{{\textbf{w}_i}^*\right\}_{i= 1}^{K}\) is the optimal solution of \({\mathcal{P}3} \). For the case of the proposed CI scheme, since the received symbols already lie in the constructive region of the constellation as shown in Fig. 2 and Fig. 3, equalization is not required by the downlink users. This automatically translates to reduced complexity at the receiver. Accordingly, this implies that CSI is not required for detection at the downlink users for the proposed CI scheme. Thus, depending on the signaling and pilots already involved for the SINR estimation, the proposed CI scheme may lead to further savings in training time and overhead. Most importantly, this makes the proposed scheme resistant to any quantization errors from the CSI acquisition at the receiver.

\section{Results} 
In this section, we investigate the performance of our proposed system through simulations. We model all channels as independent and identically distributed Rayleigh fading for both the perfect and imperfect CSI cases. Systems with QPSK and 16QAM modulation are considered while it is clear that the benefit extends to any lower or higher order modulation. For comparison in every scenario, we compare the proposed technique, constructive interference (CI) with the conventional case i.e. when all interference is treated as harmful signal \cite{sun2015multi,sun2016multi}. We use \( N \times K \times J \) to denote an FD radio BS with \( N \) antennas, \( K \) downlink users and \( J \) uplink users, respectively. \par    
\begin{figure}[t]
\centering
\includegraphics[width=8cm]{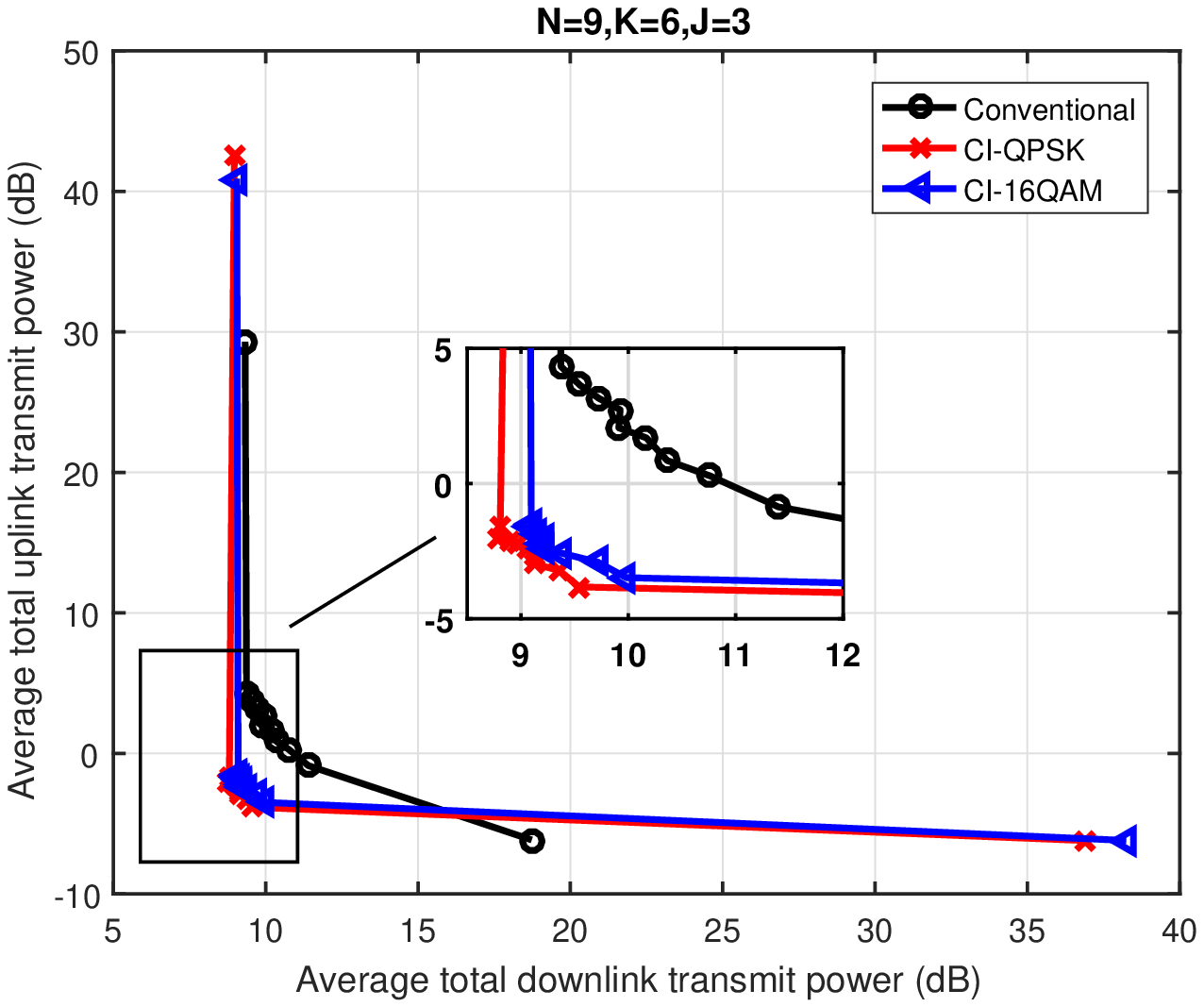}  
\caption{Average system object trade-off region achieved by the proposed scheme versus the conventional scheme N = 9, K = 6, J = 3.} 
\end{figure} 

\subsection{Uplink-Downlink Power Trade-off}
In Fig. 4, we investigate the trade-off between the downlink and uplink total transmit power for the case of \( N=9, K = 6, J=3 \) antennas. The trade-off region is obtained by solving problem \( {\mathcal{P}3} \) and \( {\mathcal{P}6} \) for the conventional and CI case, respectively, for \( 0 \leq{\lambda_{a}} \leq 1, a\in(1,2) \) with a step size of 0.1. Note that \( {\lambda_{a}} \) determines the priority of the \( \textit{a-th} \) objective. We assume the same required SINR for all downlink users to be \( \Gamma_i^{DL} = 10dB \) and \( \Gamma_i^{UL} = 0dB \) for all uplink users. It can be seen from the plot that there is a trade-off between the two objectives (downlink and uplink) i.e. an increase in one leads to a decrease in the other and vice versa. 
\begin{figure}[t]
\centering
\includegraphics[width=8cm]{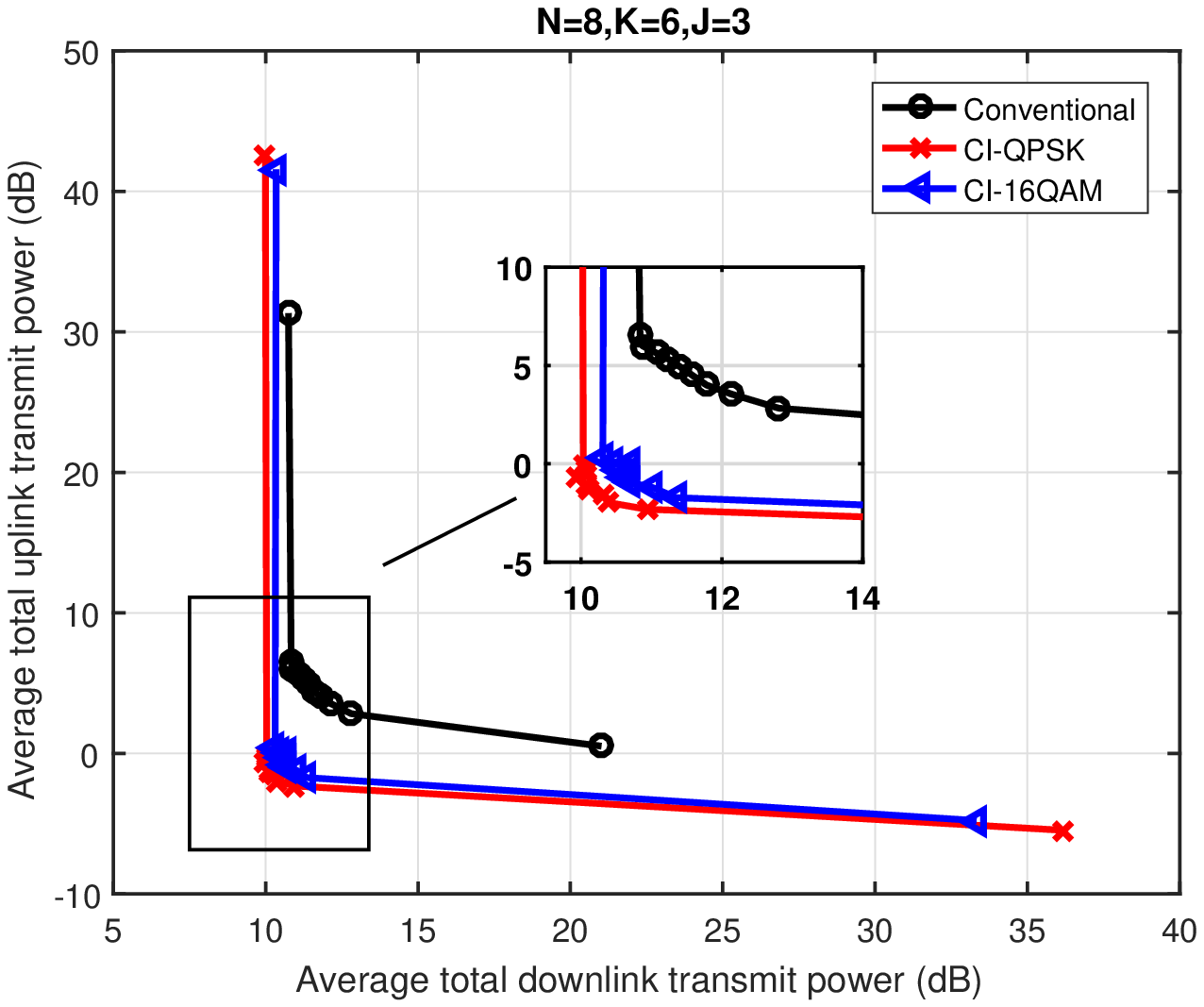}  
\caption{Average system object trade-off region achieved by the proposed scheme versus the conventional scheme N = 8, K = 6, J = 3.} 
\end{figure}
\begin{figure}[t]
\centering
\includegraphics[width=8cm]{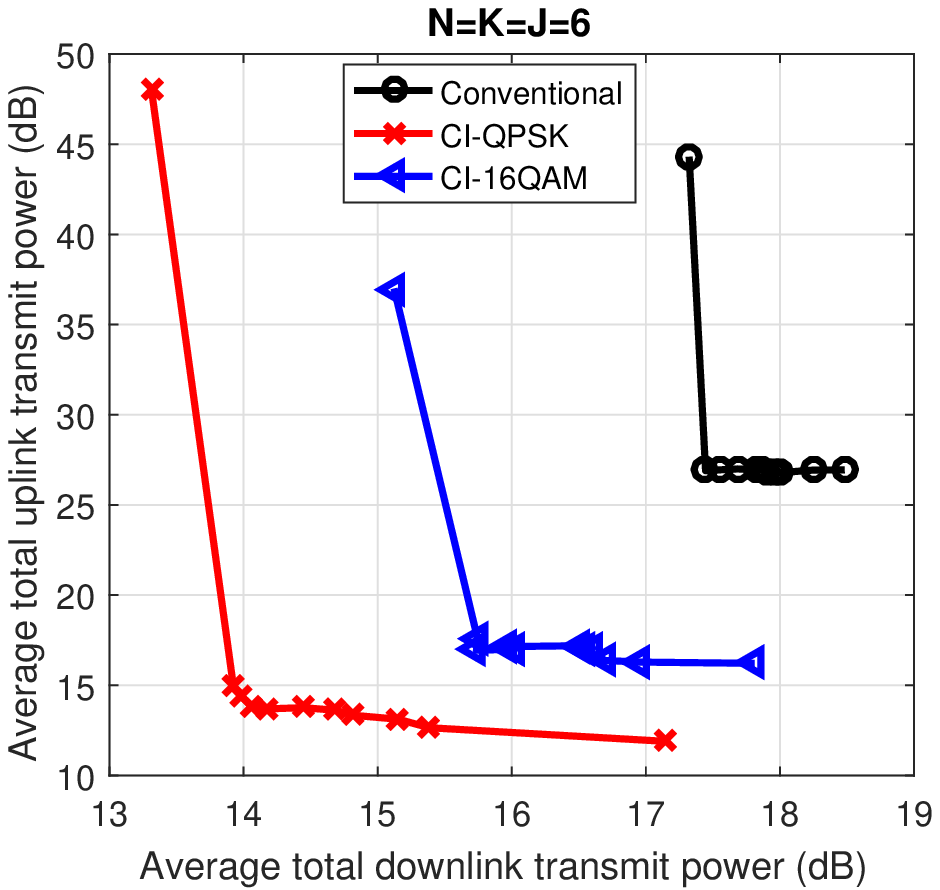}  
\caption{Average system object trade-off region achieved by the proposed scheme versus the conventional scheme N = 6, K = 6, J = 6.} 
\end{figure}
We compare the trade-off plot for the conventional scheme and the CI schemes when applied to QPSK and 16QAM modulations. We study the trade-off plots when the total number of antennas at the users is equal to the number of antennas at the FD radio BS. Thus, we can observe that the CI scheme has power savings of about 7dB for the uplink users in both QPSK and 16QAM modulations and about 2dB and 1.2dB power savings for the downlink users in QPSK and 16QAM modulations, respectively. Note that the proposed scheme is only outperformed by the conventonal beamforming for the case \(\lambda_1=0, \lambda_2=1 \), where all priority is given to the uplink PM problem, where interference exploitation does not apply.         
\par In Fig. 5, we plot the case when we have \( N=8,K=6,J=3 \). The same trend can be seen with Fig. 4, where we have for QPSK modulation power savings of about 6dB and 2dB for the uplink and downlink users, respectively. And for 16QAM modulation, we have power savings of about 6dB and 1.8dB for the uplink and downlink users, respectively. This two scenarios \( N=9, K=6, J=3 \) and \( N=8,K=6,J=3 \), show a practical perspective in the sense that there is usually more antennas at the FD radio BS than the number of antennas at the users and the optimisation problems are always feasible. 
\par In Fig. 6, we show a scenario where we have equal number of antennas at the FD radio BS and at the users \( N=K=J=6 \). With this setup we can see for QPSK modulation uplink and downlink user power savings of about 12dB and 4dB, respectively,  and about 10dB and 2dB, respectively, for 16QAM modulation. The reason is because for \( N=K=J=6 \) the problem is more restricted in the optimisation variable dimensions and the conventional scheme in this scenario leads to greatly increased uplink and downlink powers while for the CI scheme this scenario can be accommodated and has higher feasibility so consumes lower power. These results highlight a key advantage of the proposed scheme over the conventional approaches.

\begin{figure}[t]
\centering
\includegraphics[width=8cm]{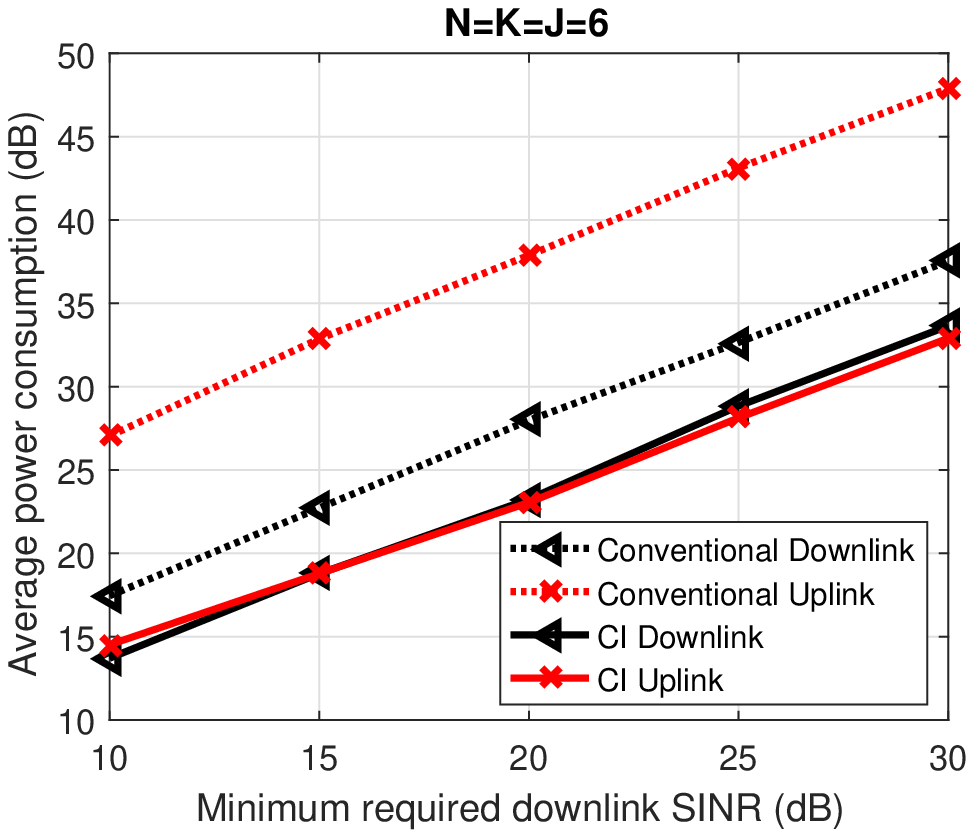}  
\caption{Average power consumption versus minimum required downlink SINR when \( \lambda_1=0.9, \lambda_2=0.1\) and \( \Gamma^{UL}=0\)dB for QPSK modulation }
\end{figure}
\begin{figure}[t]
\centering
\includegraphics[width=8cm]{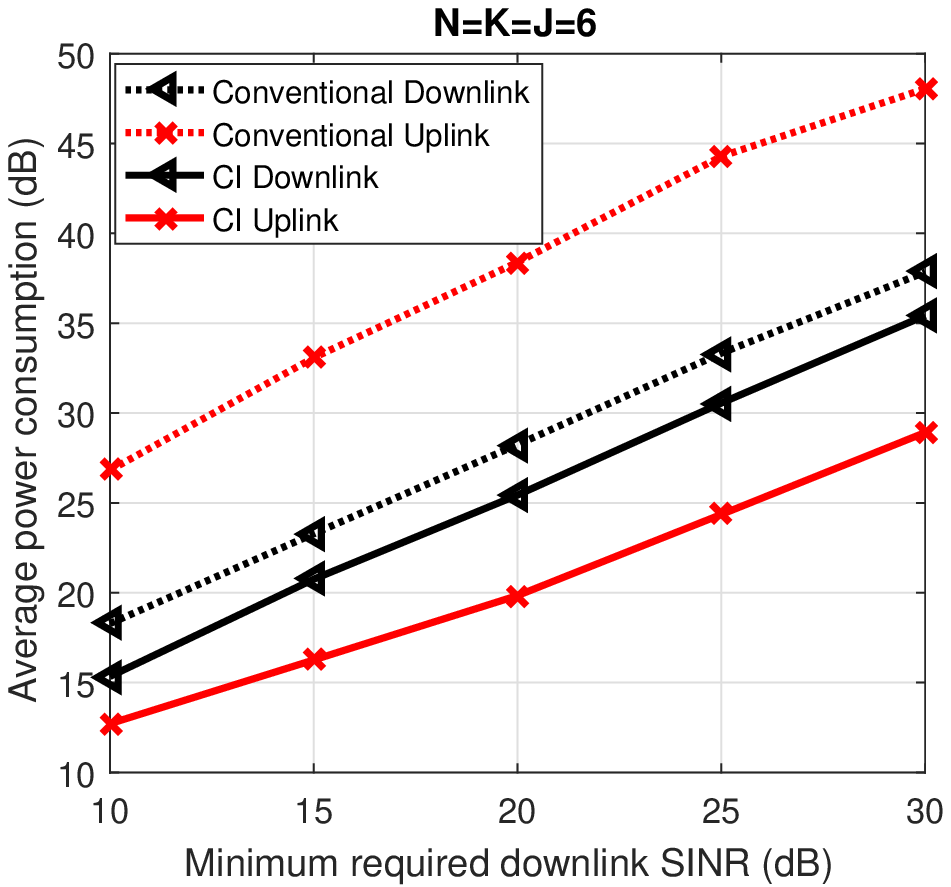}  
\caption{Average power consumption versus minimum required downlink SINR when \( \lambda_1=0.1, \lambda_2=0.9\) and \( \Gamma^{UL}=0\)dB for QPSK modulation }
\end{figure}
\begin{figure}[t]
\centering
\includegraphics[width=8cm]{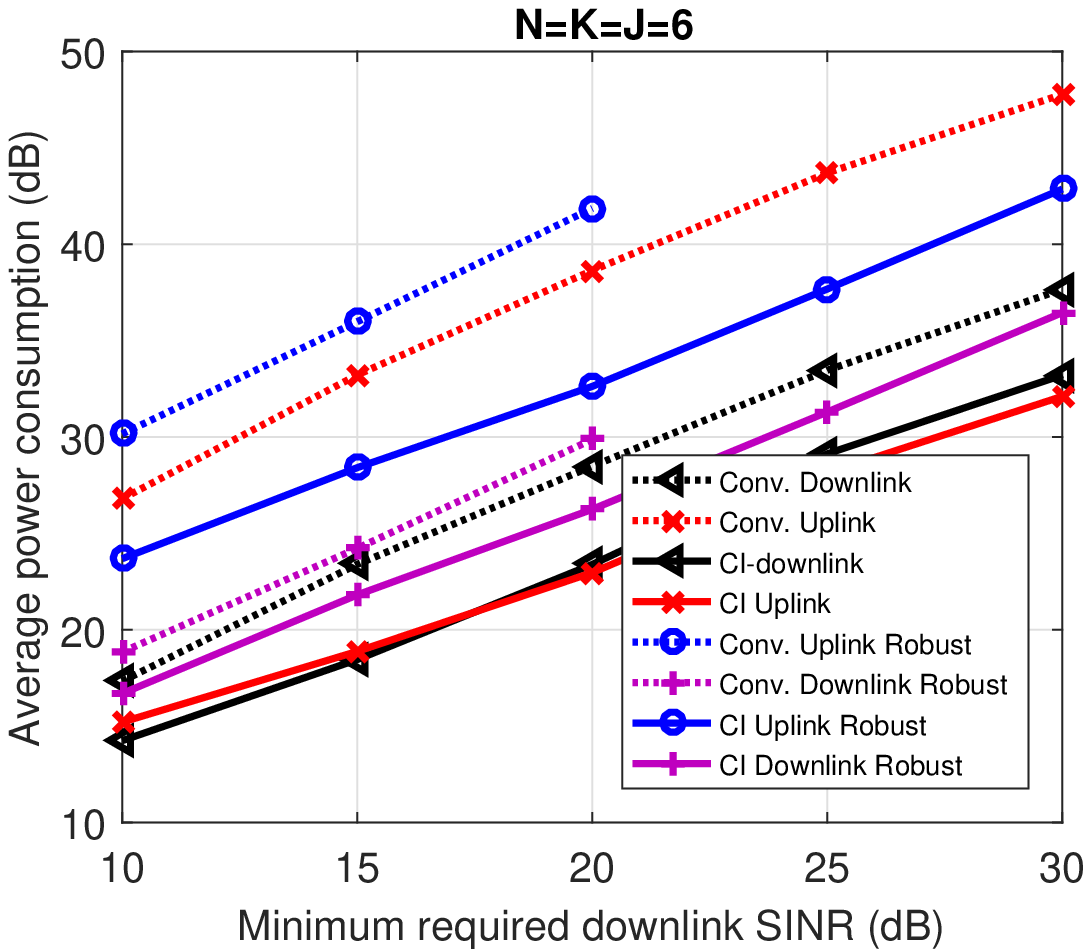}  
\caption{Average power consumption versus minimum required downlink SINR when \( \lambda_1=0.9, \lambda_2=0.1, \Gamma^{UL}=0\)dB and \( \epsilon_h = \epsilon_f = \epsilon_G = 0.1 \) for QPSK modulation } 
\end{figure}

\subsection{Average Transmit Power versus Minimum Required SINR}
In Fig. 7 and Fig. 8, we investigate the power consumption of the downlink and uplink users for different minimum required downlink SINR (\(\Gamma_i^{DL} \)). For both plots we assume a minimum required uplink SINR \(\Gamma_j^{UL} = 0dB \) for all uplink users. 
In Fig. 7, we select \( \lambda_1=0.9 \) and \( \lambda_2=0.1. \) which gives higher priority to the total downlink transmit power minimisation problem. It can be observed that both the uplink and downlink power consumption increases with increase in \(\Gamma_i^{DL} \). This is because an increase in the downlink SINR requirment translates to increace in downlink transmit power and hence increase in the SI power. Therefore, the uplink users have to transmit with a higher power to meet their QoS requirement (\(\Gamma_j^{UL} \)). However, we can still see power savings of 12dB and 5dB for the uplink and downlink users, respectively, for the CI scheme compared to the conventional scheme. Also, we note that while CI is applied to only the downlink users, more power is saved for the uplink users than the downlink users. This is because with CI the total downlink transmit power is reduced and this directly reduces the residual SI power at the FD BS. Accordingly, the constructive interference power has been traded off for both uplink and downlink power savings. The same trend can be seen in the Fig. 8, where \( \lambda_1=0.1 \) and \( \lambda_2=0.9 \). It can be observed that in this scenario since we give higher priority to the uplink power minimisation problem, we have higher power savings for the uplink users and lower power savings for the downlink users compared to the Fig. 7.
\subsection {MOOP with Imperfect CSI}
In Fig. 9 and 10, we investigate the performance of the proposed CSI-robust CI scheme for \(N=K=J=6\), we select \( \lambda_1=0.9 \) and \( \lambda_2=0.1 \). 
\begin{figure}[t]
\centering
\includegraphics[width=8cm]{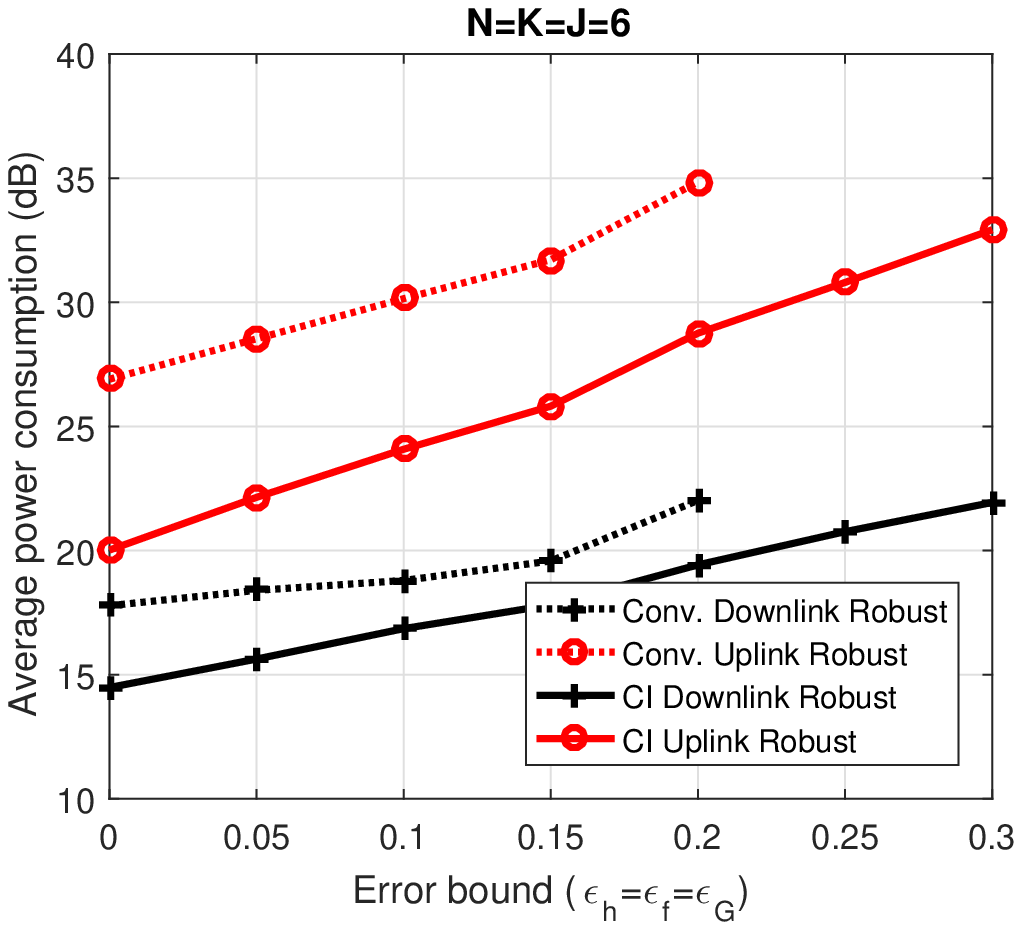}  
\caption{Average power consumption versus error bounds when \( \lambda_1=0.9, \lambda_2=0.1, \Gamma^{UL} = 0dB \) and \( \Gamma^{DL} = 10dB  \) for QPSK modulation }
\end{figure}
\begin{figure}[t]
\centering
\includegraphics[width=8cm]{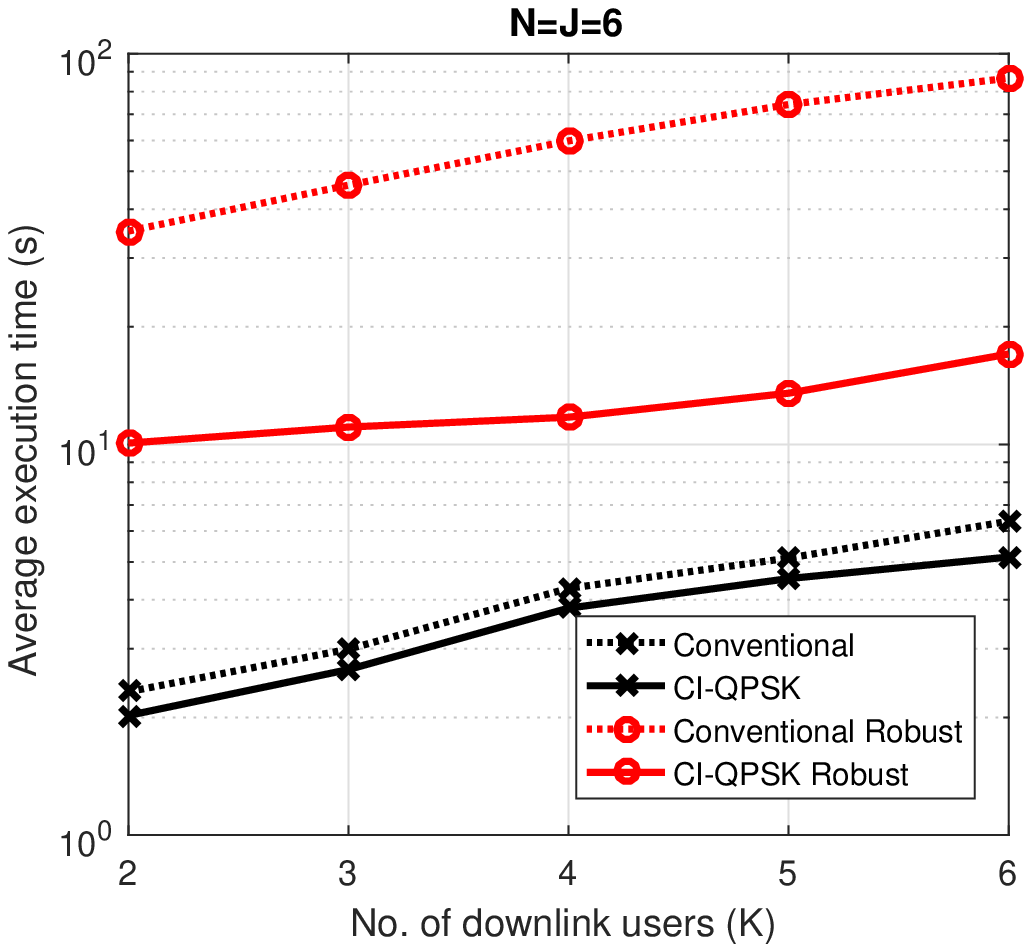}  
\caption{Average execution time per optimisation versus number of downlink users with \( N=J=6 \) when \( \lambda_1=0.9, \lambda_2=0.1, \Gamma^{UL}=0\)dB, \( \Gamma^{DL}=5\)dB and \( \epsilon_h = \epsilon_f = \epsilon_G = 0.01 \)  } 
\end{figure}
Fig. 9 shows the Average power consumption for the uplink and downlink users when the error bounds \( \epsilon_h=\epsilon_f=\epsilon_G = 0.1 \). It can be seen that the CI scheme shows better performance than the conventional scheme with power savings of 6dB and 4dB for the uplink and downlink users, respectively. In addition, for the conventional cases, feasible solutions can only be found for minimum required downlink SINR \(\Gamma_i^{DL}\leq20\)dB. This indicates that the channel error tolerance of the conventional scheme is much lower than that of the proposed CI scheme. This is also shown in Fig. 10, which shows the average power consumption with increasing error bounds. It can be seen that feasible solutions can only be found for \( \epsilon_h=\epsilon_f=\epsilon_G \leq 0.2 \). Besides, even if feasible results could be found, significant amount of power will be consumed as can be seen for error bound values between 0.15 and 0.2 for both uplink and downlink users.         

\subsection{Complexity }
In Fig. 11, we compare the Average execution time per optimisation of the conventional scheme and the proposed CI scheme for different number of downlink users (\(K\)) with \( N=J=6 \). We fixed \( \lambda_1=0.9, \lambda_2=0.1, \Gamma^{UL}=0\)dB, \( \Gamma^{DL}=5\)dB and \( \epsilon_h = \epsilon_f = \epsilon_G = 0.01. \) It can be seen that for the perfect CSI case, the proposed CI scheme takes 83\% of time taken by the conventional scheme. While for the imperfect CSI case, the proposed CI scheme takes about 28\% of the time taken by the conventional scheme. This is because the conventional approach involves a more complicated set of constraints, hence, more computational cost as shown in Section VI-A above. Besides, the proposed MOOP \( {\mathcal{P}14}\) formulation involves a multicast approach which reduces the number variables to compute. 
\par As we have noted above however, the proposed data dependent optimization needs to be run on a symbol-by-symbol basis. To obtain a fairer comparison, we plot in Fig. 12 the average execution time per frame versus the number of downlink users for slow and fast fading channels. Here, we assume the LTE Type 2 TDD frame structure \cite{LTE2008}, where each frame is subdivided to 10 subframes each with a duration \(1ms\) and containing 14 symbol-time slots. Accordingly, we assume that for fast fading the channel is constant for the duration of a subframe with a number of symbols per  coherence time \(N_{coh}=14\), while for slow fading we assume a coherence time equal to 5 subframes with \(N_{coh}=70\) \cite{LTE2008}. The results for both slow and fast fading channels show the end complexity of the proposed CI approaches are comparable to those with the conventional approaches. Accordingly, and in conjunction with the performance improvements shown in the previous results, it can be seen that the proposed schemes provide a much more favorable performance complexity trade-off w.r.t. conventional interference mitigation.         

\begin{figure}[t]
\centering
\includegraphics[width=8cm]{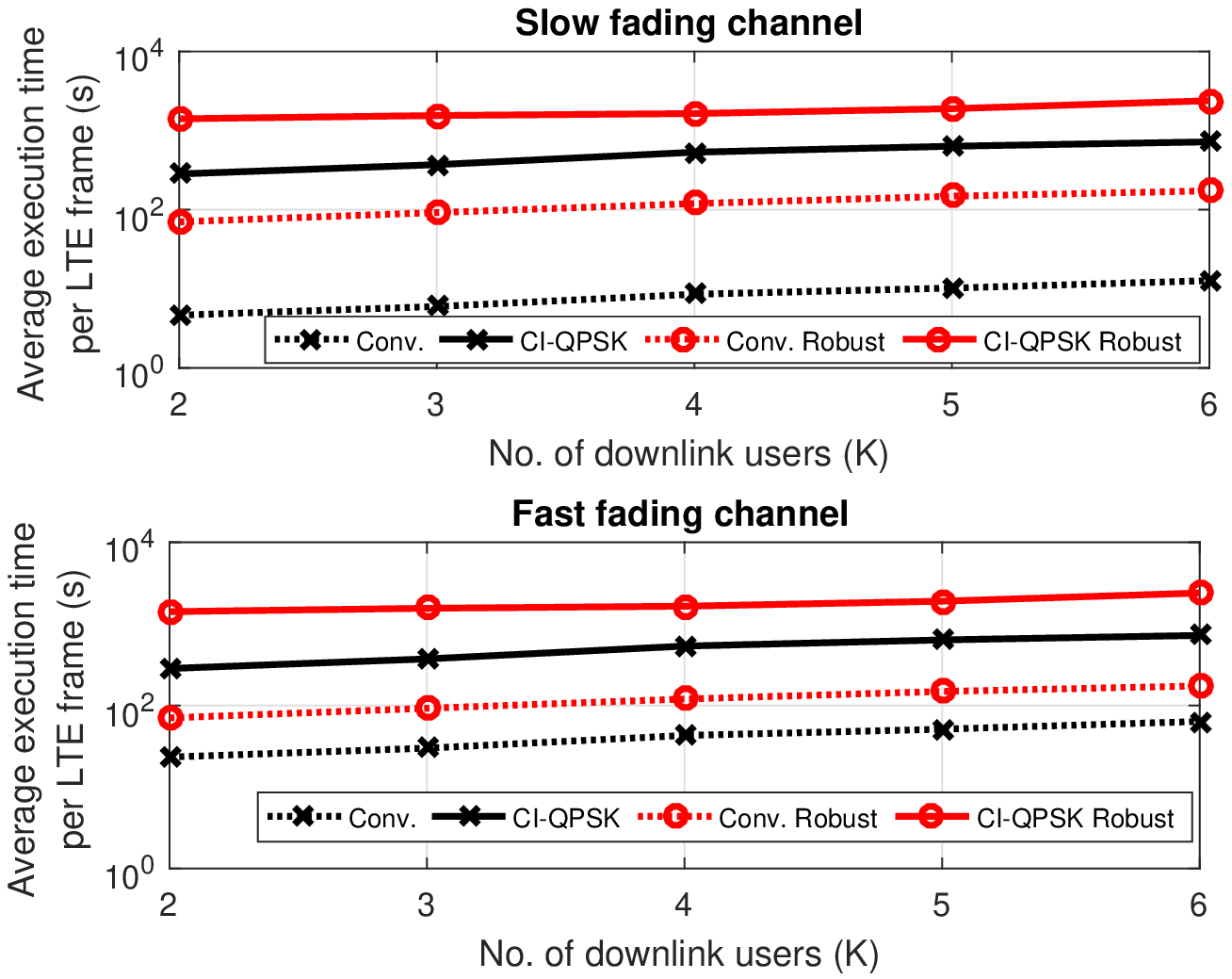}  
\caption{Average execution time versus number of downlink users for slow/fast fading channels with \( N=J=6 \) when \( \lambda_1=0.9, \lambda_2=0.1, \Gamma^{UL}=0\)dB, \( \Gamma^{DL}=5\)dB and \( \epsilon_h = \epsilon_f = \epsilon_G = 0.01. \) } 
\end{figure}

\section{Conclusion}
In this paper we studied the application of the interference exploitation concept to a MU-MIMO system with a FD radio BS. The optimisation problem was formulated as a convex Multi-Objective Optimisation problem (MOOP) via the weighted Tchebycheff method. The MOOP was formulated for both PSK and QAM modulations by adapting the decision thresholds in both cases to accommodate for constructive interference. The CI scheme was also extended to robust designs for imperfect downlink, uplink and SI CSI with bounded CSI errors. Simulation results proved the significant power savings of the CI scheme over the conventional scheme in every scenario. More importantly, we have shown that through the FD MOOP formulation, constructive interference power can be traded off for both uplink and downlink power savings.         
\section*{Acknowledgment}
The author would like to thank the Federal Republic of Nigeria and the Petroleum Technology Development Fund (PTDF) for funding his PhD.



\bibliographystyle{IEEEtran}
\bibliography{ref}
\end{document}